\documentclass[11pt]{article}

\usepackage[preprint]{acl}
\usepackage{times}
\usepackage{latexsym}
\usepackage[T1]{fontenc}
\usepackage[utf8]{inputenc}
\usepackage{microtype}
\makeatletter
\def\MASPRM@patch@showhyphens#1[#2]#3\DeclareRobustCommand\showhyphens[#4]#5{%
  \patchcmd#1{\color@begingroup}{\color@begingroup\pdfadjustspacing\z@}{}%
    {\DeclareRobustCommand\showhyphens[#4]{#5}}}
\patchcmd{\MT@setup@expansion}{\CheckCommand}{\MASPRM@patch@showhyphens}{}{}
\makeatother
\usepackage{inconsolata}
\usepackage{graphicx}
\usepackage{booktabs}
\usepackage{amsmath,amssymb}
\usepackage{multirow}
\usepackage{array}
\usepackage{enumitem}

\usepackage{hyperref}
\usepackage[nameinlink,capitalise]{cleveref}
\usepackage{placeins}
\usepackage{dblfloatfix}
\usepackage{flafter}

\usepackage{algorithm}
\usepackage{algpseudocode}
\algrenewcommand\algorithmicrequire{Input:}
\algrenewcommand\algorithmicensure{Output:}

\newlength{\MASPRMorigTextFloatSep}
\newlength{\MASPRMorigInTextSep}
\newlength{\MASPRMorigFloatSep}
\newlength{\MASPRMorigDblTextFloatSep}
\newlength{\MASPRMorigDblFloatSep}
\newlength{\MASPRMorigAboveDisplaySkip}
\newlength{\MASPRMorigBelowDisplaySkip}
\newlength{\MASPRMorigAboveDisplayShortSkip}
\newlength{\MASPRMorigBelowDisplayShortSkip}
\newlength{\MASPRMorigJot}
\newlength{\MASPRMorigAboveCaptionSkip}
\newlength{\MASPRMorigBelowCaptionSkip}
\newcommand{\MASPRMsaveSpacing}{%
  \setlength{\MASPRMorigTextFloatSep}{\textfloatsep}%
  \setlength{\MASPRMorigInTextSep}{\intextsep}%
  \setlength{\MASPRMorigFloatSep}{\floatsep}%
  \setlength{\MASPRMorigDblTextFloatSep}{\dbltextfloatsep}%
  \setlength{\MASPRMorigDblFloatSep}{\dblfloatsep}%
  \setlength{\MASPRMorigAboveDisplaySkip}{\abovedisplayskip}%
  \setlength{\MASPRMorigBelowDisplaySkip}{\belowdisplayskip}%
  \setlength{\MASPRMorigAboveDisplayShortSkip}{\abovedisplayshortskip}%
  \setlength{\MASPRMorigBelowDisplayShortSkip}{\belowdisplayshortskip}%
  \setlength{\MASPRMorigJot}{\jot}%
  \setlength{\MASPRMorigAboveCaptionSkip}{\abovecaptionskip}%
  \setlength{\MASPRMorigBelowCaptionSkip}{\belowcaptionskip}%
}
\newcommand{\MASPRMcompactMainSpacing}{%
  \setlength{\textfloatsep}{4pt plus 1pt minus 2pt}%
  \setlength{\intextsep}{5pt plus 1pt minus 2pt}%
  \setlength{\floatsep}{4pt plus 1pt minus 1pt}%
  \setlength{\dbltextfloatsep}{4pt plus 1pt minus 2pt}%
  \setlength{\dblfloatsep}{4pt plus 1pt minus 1pt}%
  \setlength{\abovedisplayskip}{6pt plus 2pt minus 3pt}%
  \setlength{\belowdisplayskip}{6pt plus 2pt minus 3pt}%
  \setlength{\abovedisplayshortskip}{0pt plus 2pt}%
  \setlength{\belowdisplayshortskip}{3pt plus 2pt minus 2pt}%
  \setlength{\jot}{2pt}%
  \setlength{\abovecaptionskip}{3pt plus 1pt minus 1pt}%
  \setlength{\belowcaptionskip}{0pt plus 1pt}%
}
\newcommand{\MASPRMrestoreSpacing}{%
  \setlength{\textfloatsep}{\MASPRMorigTextFloatSep}%
  \setlength{\intextsep}{\MASPRMorigInTextSep}%
  \setlength{\floatsep}{\MASPRMorigFloatSep}%
  \setlength{\dbltextfloatsep}{\MASPRMorigDblTextFloatSep}%
  \setlength{\dblfloatsep}{\MASPRMorigDblFloatSep}%
  \setlength{\abovedisplayskip}{\MASPRMorigAboveDisplaySkip}%
  \setlength{\belowdisplayskip}{\MASPRMorigBelowDisplaySkip}%
  \setlength{\abovedisplayshortskip}{\MASPRMorigAboveDisplayShortSkip}%
  \setlength{\belowdisplayshortskip}{\MASPRMorigBelowDisplayShortSkip}%
  \setlength{\jot}{\MASPRMorigJot}%
  \setlength{\abovecaptionskip}{\MASPRMorigAboveCaptionSkip}%
  \setlength{\belowcaptionskip}{\MASPRMorigBelowCaptionSkip}%
}
\AtBeginDocument{\MASPRMsaveSpacing\MASPRMcompactMainSpacing}

\usepackage{mathtools}
\usepackage[most]{tcolorbox}

\usepackage{listings}
\usepackage{xcolor}
\usepackage{colortbl}
\definecolor{MASPRMRowBlue}{HTML}{EAF3FF}

\usepackage[normalem]{ulem} 

\author{
  Milad Yazdani\textsuperscript{1},
  Mahdi Mostajabdaveh\textsuperscript{2}\thanks{Corresponding Author},
  Zirui Zhou\textsuperscript{2},
  Ying Xiong\textsuperscript{2}
\\
\\
  \textsuperscript{1}\small Department of Electrical and Computer Engineering, University of British Columbia, Vancouver, BC V6T1Z4, Canada \\
  \textsuperscript{2}\small Huawei Technologies Canada, Burnaby, BC V5C6S7, Canada
\\
  \small{
    Correspondence: \href{mailto:mahdi.mostajabdaveh1@huawei.com}{mahdi.mostajabdaveh1@huawei.com}
  }
}

\title{
MASPRM: Multi-Agent System Process Reward Model
}
\begin{document}
\maketitle
\begin{abstract}
Inference-time search over multi-agent systems (MAS) wastes compute when it cannot identify which agent's intermediate message advanced progress. We present the Multi-Agent System Process Reward Model (\textsc{MASPRM}), which scores routed transcripts (ordered sequences of messages between agents) and acts as an inference controller for step-level beam search (SBS) and Monte Carlo Tree Search (MCTS). \textsc{MASPRM} is trained from multi-agent MCTS rollouts labeled only with terminal outcome rewards, without human step-level annotations. We evaluate on \textsc{GSM8K}, \textsc{MATH}, \textsc{MMLU}, and \textsc{LogiQA}.
Under matched scorer size and comparable MCTS budget, \textsc{MASPRM} exceeds a size-matched ORM by $+2.0$ to $+3.0$ points at 1.5B and $+4.1$ to $+14.5$ at 7B across all four benchmarks, with additional scorer-scaling gains over policy likelihood at 7B (avg $+13.4$ under MCTS). \textsc{MASPRM} also improves ranking quality, reducing Hit@1 to Hit@5 gaps by up to $10.3$ points, with the largest gains under stepwise search that uses intermediate decisions. Code:
\url{https://github.com/milad1378yz/MASPRM}
\end{abstract}

\begin{figure}[!t]
\centering
\includegraphics[width=\linewidth,trim=6pt 4pt 2pt 4pt,clip]{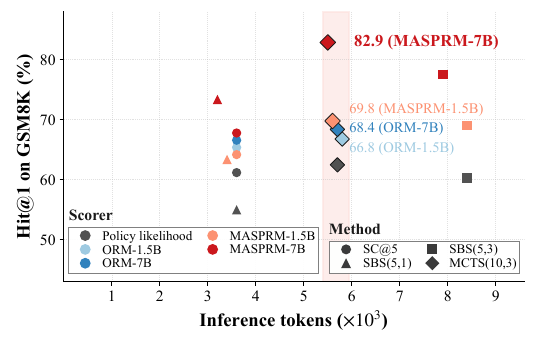}
\vspace{-8pt}
\caption{\textsc{GSM8K} Hit@1 vs.\ inference tokens. Color denotes scorer and marker denotes inference method. Under matched scorer size at comparable inference tokens, MCTS$(10,3)$+\textsc{MASPRM} exceeds ORM by $+3.0$ Hit@1 at 1.5B and $+14.5$ at 7B, reaching 82.9 with the 7B scorer.}
\label{fig:gsm8k_tradeoff}
\vspace{-4pt}
\end{figure}

\begin{figure*}[!thb]
  \centering
  \includegraphics[width=0.95\textwidth,trim=0pt 1pt 0pt 5pt,clip]{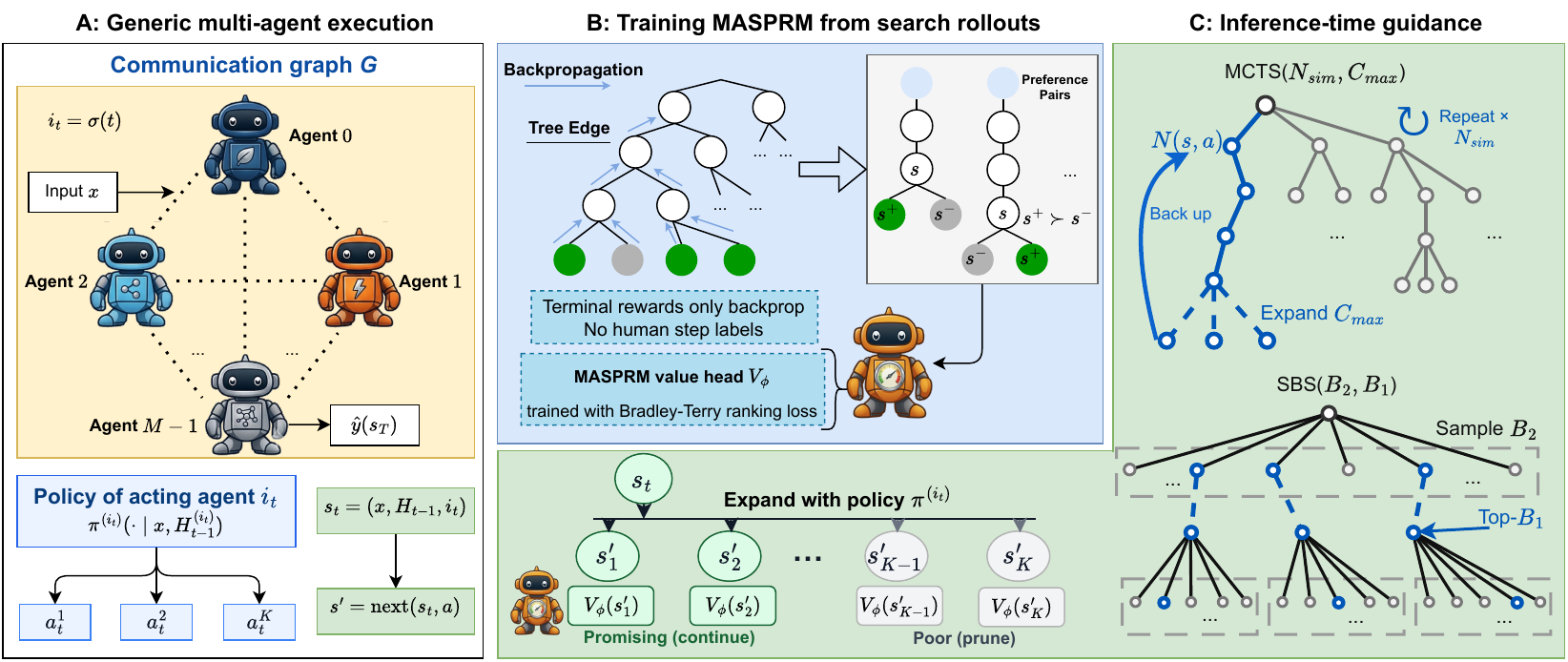}
  \vspace{-1pt}
  \caption{\textsc{MASPRM} pipeline. A: scheduled agents act over graph $G$ from local views. B: MAS-MCTS backs up terminal rewards to edges, forms sibling preferences, and trains $V_\phi$ with Bradley-Terry loss. C: MCTS and SBS expand policy-sampled successors and rank them with $V_\phi$ to continue promising branches and prune poor ones.}
  \label{fig:data}
  \vspace{-2pt}
\end{figure*}

\vspace{-6pt}
\section{Introduction}
\vspace{-6pt}
Single-pass LLM reasoning is unreliable when a problem requires exploring alternatives, justifying choices, and coordinating subtasks~\cite{wei2022chain,zhou2023leasttomost,wang2023plan,zhang2024g,zhou2026multiagent}. Multi-agent systems (MAS) address this limitation by running multiple agents that interact over a communication graph specifying who can message whom, often with distinct roles and capabilities. We call an MAS execution's ordered sequence of routed messages a routed transcript: at each step, an acting agent proposes an action and routes it to recipients. This structure naturally supports inference-time procedures that allocate compute to exploration, verification, and ranking, rather than commit to one routed transcript. Indeed, the community has found that \emph{how} compute is spent at inference can matter as much as, or more than, parameter count: expanding and scoring partial solutions yields substantial gains in correctness and confidence~\cite{snell2025scaling,lightman2023let}. This is especially relevant for smaller LLMs, whose low call cost makes multi-agent search feasible and structured exploration effective~\cite{guan2025rstar,wang2023selfconsistency,yao2023tree}. Yet realizing these gains in MAS depends on reliable guidance over long routed transcripts.

Despite this promise, inference search in MAS remains unreliable for two reasons. First, answer-level methods such as majority voting and verifier reranking help choose among completed solutions~\cite{wang2023selfconsistency,lightman2023let}, but they do not by themselves identify which agent's intermediate message advanced progress. Second, errors can accumulate across steps. Without scoring routed transcript prefixes, search may keep extending prefixes that are likely under the policy but misaligned with eventual correctness, wasting compute on unproductive directions. Prior work in single-agent settings shows that feedback on intermediate reasoning can guide inference-time search toward trajectories where improvement is more likely~\cite{setlur2025rewarding,lightman2023let,chen2024alphamath,guan2025rstar,zeng2025versaprm}. However, extending these ideas to MAS is not straightforward.

We introduce \textsc{MASPRM}, a process reward model for MAS. Unlike prior PRMs that score single-agent reasoning chains, \textsc{MASPRM} scores routed transcript prefixes and agents' proposed actions, enabling agent-aware credit assignment in collaborative settings. These scores act as a controller during inference, guiding expansion and compute allocation over candidate agent rollouts. Our main experiments study predefined MAS configurations on verifiable terminal-answer tasks.

Developing PRMs for MAS raises four challenges beyond single-agent reasoning. (1) \emph{Score granularity}: a ``step'' is an agent action (planning, routing, tool use) rather than a small reasoning act, so the PRM should score routed transcript prefixes rather than tokens. (2) \emph{Schedule and graph}: the value of a transcript depends on the order agents act and what messages route where. (3) \emph{Heterogeneity}: agents differ in role, tool access, and prompts, so intermediate prefixes come from different distributions. (4) \emph{Partial observability}: each agent observes only its local view of the global routed transcript.

To address these challenges, we let each acting agent propose candidate actions from its local view, while the \textsc{MASPRM} scorer conditions on the full routed transcript prefix. This preserves partial observability for generation but provides a consistent global routed-transcript representation for guiding inference-time search. The scorer also conditions on the acting agent's identity and the transcript ordering, so heterogeneity and the schedule are encoded in the input rather than handled by a separate mechanism. These design choices draw on lessons from structured exploration~\cite{wang2023selfconsistency,yao2023tree,snell2025scaling} and search guided by verifiers~\cite{lightman2023let}. They also remain compatible with implicit process signals learned from outcome supervision alone~\cite{cui2025process}.

\noindent\textbf{Contributions.}
\begin{itemize}[topsep=2pt,itemsep=1pt,leftmargin=*]
  \item We formulate \textsc{MASPRM}, a process reward model that scores agent actions and routed transcript prefixes.
  \item We propose MAS-MCTS supervision: backing up terminal rewards to agent-action values trains \textsc{MASPRM} without human step-level annotations.
  \item We introduce \textsc{MASPRM}-guided SBS and MCTS over agent rollouts in predefined MAS, improving decision quality and pruning unproductive branches.
\end{itemize}

Figure~\ref{fig:gsm8k_tradeoff} shows the main \textsc{GSM8K} tradeoff: under matched scorer size, MCTS$(10,3)$+\textsc{MASPRM} beats ORM by 3.0 Hit@1 at 1.5B and 14.5 at 7B, reaching 82.9 with the 7B scorer.

\vspace{-8pt}
\section{Related Work}
\vspace{-8pt}
Recent work on LLM-based MAS studies how agent roles, prompts, communication graphs, and tool-routing choices affect coordination and efficiency~\cite{zhang2024g,zhou2026multiagent}. \textsc{DyLAN} studies dynamic agent networks with diverse agents, agent communication, and adaptive scheduling~\cite{liu2024dylan}. This line is complementary to our work: rather than proposing a new communication topology or role set, our main experiments hold the MAS configuration fixed and study how to allocate inference-time computation over candidate routed transcripts.

Self-consistency, Tree-of-Thought, verifier-guided decoding, and test-time compute scaling show exploration and selection improve LLM reasoning~\cite{wang2023selfconsistency,yao2023tree,lightman2023let,snell2025scaling}. Search has also been studied in multi-agent settings. For example, \textsc{MASTER} coordinates agent recruitment and communication using an LLM-specialized MCTS procedure~\cite{gan2025master}, and \textsc{PlanGEN} integrates inference-time algorithms such as Best-of-$N$, Tree-of-Thought, and REBASE with specialized agents for planning and reasoning~\cite{parmar2025plangen}. These works show that search and adaptive inference are useful for MAS, but both \textsc{MASTER} and \textsc{PlanGEN} target orchestration over varying agent populations; \textsc{MASPRM} instead holds the MAS fixed and learns a process value model over routed transcript prefixes to guide stepwise expansion and pruning, so the methods are not directly compatible as baselines.

PRMs provide dense feedback over intermediate reasoning and have been used to guide search and prune unpromising branches, primarily in single-agent reasoning chains~\cite{setlur2025rewarding,lightman2023let}. To reduce labeling cost, several works derive process supervision from search traces. \textsc{AlphaMath Almost Zero} uses MCTS rollouts to construct step-level targets~\cite{chen2024alphamath}, \textsc{rStar-Math} pairs PRMs with MCTS for math reasoning~\cite{guan2025rstar}, and \textsc{Math-Shepherd} constructs step-level supervision for math without human annotations~\cite{wang2024mathshepherd}. More general PRMs such as \textsc{VersaPRM} study cross-domain usefulness~\cite{zeng2025versaprm}, while \textsc{ProcessBench} evaluates whether process feedback can localize reasoning errors~\cite{zheng2025processbench}. These methods are largely developed for single-agent chains where a single linear prefix defines the intermediate state. In MAS, messages are routed and agents act from different local views, so evaluating a partial state must condition on both the routed transcript and the acting agent.

Step-level process refinement provides intermediate feedback for LLM-agent trajectories~\cite{xiong2024watch}. \textsc{ARMAP} learns reward models from environment interaction and combines them with planning to guide autonomous agents~\cite{chen2025scaling}. LLM-judge approaches assign credit to agents using natural-language critics or decomposed team rewards~\cite{lin2025speaking,nagpal2025leveraging}. \citet{xi2026agentprm} proposes \textsc{AgentPRM} for process-reward learning in single-agent environment-interaction settings. \textsc{MASPRM} instead targets multi-agent inference-time search: it scores routed transcripts of communications between heterogeneous agents and is used as a search controller during SBS and MCTS, not as a training signal for the policy.

\vspace{-8pt}
\section{Preliminaries}
\vspace{-6pt}
\label{sec:preliminaries}

Each instance consists of a question $x$ and a ground truth answer $y^\star$.
We model an MAS execution as a finite routed transcript, i.e., an ordered list of routed messages along the communication graph, ending when a message is routed to the terminal node. Correctness is assessed only at termination.
We model the system as a directed graph
\(
G = \bigl(\mathcal{V}\cup\{\text{q},\text{sink}\},E\bigr),
\quad \mathcal{V} = \{0,\dots,M{-}1\},
\)
where each $i\in\mathcal{V}$ is an agent, $\text{q}$ is a non-acting question node that holds $x$, and $\text{sink}$ is a non-acting terminal node with no outgoing edges.
An edge $(i\!\to\! j)\in E$ means $i$ routes a message to $j$.
For $u\in \mathcal{V}\cup\{\text{q},\text{sink}\}$ define
\(
\mathcal{N}_{\mathrm{in}}(u)=\{\,v:(v\!\to\!u)\in E\,\}
\)
and
\(
\mathcal{N}_{\mathrm{out}}(u)=\{\,w:(u\!\to\!w)\in E\,\}.
\)

We use a fixed schedule $\sigma:\{1,\dots,D\}\to \mathcal{V}$. At turn $t\in\{1,\dots,D\}$,
if the rollout has not yet terminated, the acting agent is $i_t=\sigma(t)$. Rollouts run for at most
$D$ turns and terminate once a message is routed to $\text{sink}$. If the rollout has not terminated earlier, the final scheduled action at $t=D$ routes to $\text{sink}$. We use pre-turn indexing and let $T\le D{+}1$ denote the terminal index such that $s_T$ is terminal.
In our experimental MAS configurations, termination occurs immediately after the final scheduled turn ($T=D{+}1$),
though earlier stopping is allowed.
This formalization matches the main experiments, while App.~\ref{app:dylan-dynamic} considers a separate conditional scheduler variant.

Let $H_{t-1}$ be the routed transcript after $t{-}1$ turns, an ordered list of triples
$(\text{speaker},\text{message},\text{recipient})$ consistent with $G$, with $H_0=\varnothing$.
We define the state space $\mathcal{S}$ and represent a state as
\(
s_t = \big(x,\,H_{t-1},\,i_t\big)\in\mathcal{S}.
\)
The initial search state is \(s_1(x)=(x,\emptyset,\sigma(1))\).
Because routing limits information flow, agents do not observe the full routed transcript $H_{t-1}$. For each agent
$i\in\mathcal{V}$, let $H_{t-1}^{(i)}$ denote its local view: the ordered subsequence of $H_{t-1}$
containing only tuples where $i$ is the speaker or recipient.
In what follows, the acting agent conditions action proposals on $(x,H_{t-1}^{(i_t)},i_t)$, while the default \textsc{MASPRM} scorer conditions on $(x,H_{t-1},i_t)$. Thus, \textsc{MASPRM} has access to the full routed transcript prefix, while agent policies operate under partial observability through their local views $H_{t-1}^{(i)}$.
In the local-observability ablation, the scorer input is $H_{t-1}^{(i_t)}$ plus the acting agent's new output, rather than $H_{t-1}$.
At $s_t$, agent $i_t$ chooses an action $a_t\in\mathcal{A}_{i_t}(s_t)$, a (message, recipient) pair
where the recipient must be in $\mathcal{N}_{\mathrm{out}}(i_t)$. Here $\mathcal{A}_i(s)$ is the set of valid
actions for agent $i$ at state $s$, yielding the deterministic successor
\(
s_{t+1}=\mathrm{next}(s_t,a_t).
\)
Routing affects which agents observe messages, but turn-taking is fixed by $\sigma$. If the recipient is not $\text{sink}$, then $i_{t+1}=\sigma(t{+}1)$. Otherwise, $i_{t+1}=\text{sink}$, $s_{t+1}$ is terminal, and the rollout ends at $T=t{+}1$.

For a terminal state $s_T$, let $\hat y(s_T)$ be the extracted final answer. We define $R(s_T)=+1$ iff $\hat y(s_T)=y^\star$, and $R(s_T)=-1$ otherwise.
For any non-terminal state $s$, set $R(s)=0$.
For any state $s$ and valid action $a$,
\(
V(s)   = \mathbb{E}\!\left[R(s_T)\mid s\right], \;
Q(s,a) = \mathbb{E}\!\left[R(s_T)\mid s,a\right],
\)
where the expectation is over future actions under the rollout distribution induced by the search procedure. During training, these actions are sampled via the search policy (UCT selection at expanded nodes and policy sampling at frontiers). At inference, the learned process value head guides the improved search behavior.
Here $V(s)$ and $Q(s,a)$ denote the true population values under this rollout distribution.
Under deterministic transitions and terminal rewards,
\(
Q(s,a)=V\!\big(\mathrm{next}(s,a)\big).
\)

Each agent $i$ is paired with a policy $\pi^{(i)}(\cdot\mid x,H_{t-1}^{(i)})$ over actions,
conditioned only on its local view.
During search, we form a finite candidate set
$C_t\subset\mathcal{A}_{i_t}(s_t)$ by sampling from $\pi^{(i_t)}(\cdot\mid x,H_{t-1}^{(i_t)})$.
As a baseline scorer, we use policy likelihood guidance based on a length-normalized log-likelihood of each candidate action under the acting agent (App.~\ref{app:policy-likelihood}).
A process value head $V_\phi:\mathcal{S}\to\mathbb{R}$ is the learned \textsc{MASPRM} value estimate for any state, whether intermediate or terminal. Because $V_\phi$ is defined over states $s_t=(x,H_{t-1},i_t)$, each score is tied to the acting agent at that step. At inference, \textsc{MASPRM} scores the action $a$ by the learned estimate
\(
V_\phi\!\big(\mathrm{next}(s,a)\big).
\)

For each visited edge $(s,a)$ we maintain a visit count $N(s,a)\!\in\!\mathbb{N}$ and a
running value sum $W(s,a)\!\in\!\mathbb{R}$. The MCTS tree estimate backed up from simulations is
\(
\widehat Q(s,a)\;=\;W(s,a)/N(s,a).
\)
All successors generated during search must respect $E$
(equivalently, $\mathcal{N}_{\mathrm{out}}$) and the schedule $\sigma$.
For search hyperparameters, we use $N_{\mathrm{sim}}$ simulations per instance and $C_{\max}$ as the per-node candidate cap. Rollouts have maximum depth $D$, a single pass through the MAS schedule.

\vspace{-6pt}
\section{\textsc{MASPRM}}
\vspace{-6pt}
\label{sec:method}

\textsc{MASPRM} is a process reward model that assigns values to intermediate MAS states via a shared head $V_\phi$. Fig.~\ref{fig:data} summarizes the pipeline: agents generate candidate actions under the communication graph, MAS-MCTS converts terminal rewards into preference pairs for training, and the lower inference band shows the shared policy expansion and value scoring used by both MCTS and SBS. We train $V_\phi$ from MAS-MCTS rollouts by backing up terminal rewards into action values $\widehat Q(s,a)$ and using them as targets for successor states with no manual annotations. At inference, \textsc{MASPRM} guides MCTS via upper confidence bounds applied to trees (UCT) selection (Eq.~\eqref{eq:uct}) and guides SBS by value-based ranking of successors.

\vspace{-4pt}
\subsection{\textsc{MASPRM} training}
\vspace{-2pt}
\label{sec:masprm-training}

Selection uses UCT~\cite{kocsis2006bandit} with constant $c_{\textsc{uct}}>0$.
Let $\mathcal{A}^{\mathrm{exp}}_{i_t}(s_t)=\{\,a\in\mathcal{A}_{i_t}(s_t): N(s_t,a)>0\,\}$ denote the expanded actions at $s_t$.
\vspace{-10pt}
\small
\begin{equation}
\label{eq:uct}
\begin{aligned}
a^\star
&=\arg\max_{a\in \mathcal{A}^{\mathrm{exp}}_{i_t}(s_t)}
\Biggl\{\,\widehat Q(s_t,a) \\
&\qquad + c_{\textsc{uct}}
\sqrt{\frac{\ln\!\bigl(1+\sum_{b\in \mathcal{A}^{\mathrm{exp}}_{i_t}(s_t)}N(s_t,b)\bigr)}{1+N(s_t,a)}}\,\Biggr\}
\end{aligned}
\end{equation}
\normalsize
Unexpanded actions are introduced during expansion via candidate sampling.
We generate supervision by running MAS-MCTS under graph constraints $G$ and fixed schedule
$\sigma$. At a non-terminal state $s_t$, expansion samples a candidate set
$C_t\subset\mathcal{A}_{i_t}(s_t)$ from $\pi^{(i_t)}(\cdot\mid x,H_{t-1}^{(i_t)})$ and creates successors
$s'=\mathrm{next}(s_t,a)$ for $a\in C_t$. Simulations are evaluated only at terminal leaves,
using $v_{\mathrm{leaf}}(s_T)=R(s_T)\in\{-1,+1\}$, and then backed up to update $(N,W,\widehat Q)$. We omit
virtual visit initialization during training to avoid biasing the Monte Carlo targets. A
summary appears in Alg.~\ref{alg:train-mcts} (App.~\ref{app:algorithms}).

After rollouts, for each visited parent state $s_t$ and each expanded
action $a\in C_t$ with $N(s_t,a)>0$ and child $s'=\mathrm{next}(s_t,a)$, we define a process target
\(
y_{\mathrm{proc}}(s') \;\triangleq\; \widehat Q(s_t,a)\;\in[-1,1],
\)
a Monte Carlo estimate of $Q(s_t,a)$ backed up from terminal rewards.
Collecting these labeled post action states yields the process dataset
\(
\mathcal{D}_{\mathrm{proc}}=\{(s',y_{\mathrm{proc}}(s'))\}.
\)
We then construct a preference pair dataset by comparing sibling successors within each expanded parent state. Let
$
\mathrm{Succ}(s_t)=\{\,s'=\mathrm{next}(s_t,a)\,:\, a\in C_t \text{ and } N(s_t,a)>0\,\}
$
denote the set of expanded children of $s_t$. We create preference pairs
\vspace{-4pt}
\begin{align}
\mathcal{D}_{\mathrm{pref}}
=\bigcup_{s_t} \, \{\,(s^+,s^-) |\, s^+,s^-\in \mathrm{Succ}(s_t),\nonumber\\
y_{\mathrm{proc}}(s^+)>y_{\mathrm{proc}}(s^-) \,\}
\label{eq:d_pref}
\end{align}
\vspace{-8pt}

We train $V_\phi$ using a Bradley-Terry ranking loss~\cite{ouyang2022training}. We use ranking rather than pointwise regression because at inference the scorer is used to rank candidate successor states during search.
\vspace{-10pt}
\[
p_\phi\!\left(s^+ \succ s^-\right)\;=\;\frac{\exp\!\left(V_\phi(s^+)\right)}{\exp\!\left(V_\phi(s^+)\right)+\exp\!\left(V_\phi(s^-)\right)},
\]
\vspace{-10pt}
\[
\min_{\phi}\;\frac{1}{|\mathcal{D}_{\mathrm{pref}}|}\sum_{(s^+,s^-)\in\mathcal{D}_{\mathrm{pref}}}
-\log p_\phi\!\left(s^+ \succ s^-\right).
\]
\normalsize
Because comparisons are formed among siblings, each preference pair comes from the same parent state $s_t$ and differs only in the branching action.

As a baseline that uses only terminal outcomes, we train an outcome reward model (ORM) $\Gamma$ on the same rollouts (App.~\ref{app:orm}) and use it as an outcome scorer in our experiments.

\vspace{-4pt}
\subsection{Inference guidance}
\vspace{-4pt}
\label{subsec:inference}

For inference search, we score successor states $s'=\mathrm{next}(s_t,a)$ with
$V_\phi(s')$ or a policy likelihood score $\tilde\psi_{\text{pol}}(a\mid s_t)$ (App.~\ref{app:policy-likelihood}). In MCTS, newly
created edges may be initialized with a virtual visit by setting $N(s_t,a)=1$ and
 $W(s_t,a)=\widehat Q(s_t,a)=f_{\textsc{init}}(s_t,a)$, where $f_{\textsc{init}}(s_t,a)=V_\phi(s')$
(\textsc{MASPRM}) or $f_{\textsc{init}}(s_t,a)=\tilde\psi_{\text{pol}}(a\mid s_t)$.
At inference, MCTS uses UCT selection (Eq.~\eqref{eq:uct}). At a frontier non-terminal state $s_t$,
we sample $C_t$ from $\pi^{(i_t)}(\cdot\mid x,H_{t-1}^{(i_t)})$ and create successors $s'=\mathrm{next}(s_t,a)$ for
$a\in C_t$ (with virtual visit initialization as above). Simulation leaves are evaluated by
bootstrap $v_{\mathrm{leaf}}(s)=V_\phi(s)$ (App.~\ref{app:policy-likelihood} gives the policy likelihood definition), and values are backed up to update $(N,W,\widehat Q)$. We decode
by greedily following the child with maximal $\widehat Q$ from the initial state \(s_1(x)\). Alg.~\ref{alg:infer-mcts}
(App.~\ref{app:algorithms}) summarizes the procedure.

For SBS, at depth $t$, we expand each beam state $s\in\mathcal{B}_t$ by sampling $B_2$ actions and scoring each successor by the chosen scorer. Here
$\mathcal{B}_t$ is the beam at depth $t$, $B_2$ is the number of sampled actions per beam state, and
$B_1$ is the beam width. We pool all $|\mathcal{B}_t|\,B_2$ successors and keep the top-$B_1$ to form
$\mathcal{B}_{t+1}$, repeating for $t=1,\dots,D$, and return the highest scoring terminal state in the final beam.
Alg.~\ref{alg:sbs} (App.~\ref{app:algorithms}) summarizes SBS.

\vspace{-6pt}
\section{Experiments \& Results}
\vspace{-6pt}
\label{sec:exp_res}

We design our experiments around three questions. \textbf{RQ1:} Does \textsc{MASPRM} improve Hit@1 over policy likelihood under matched within-method agent-call and token budgets, especially for SBS and MCTS? \textbf{RQ2:} How does \textsc{MASPRM} guidance compare to a conventional ORM under matched search method and, where applicable, matched scorer size? \textbf{RQ3:} Does a \textsc{MASPRM} trained on one dataset/agent transfer zero-shot to other datasets/agents and improve Hit@1 under matched within-method budgets? We also analyze Hit@$k$ sensitivity to illustrate the robustness of \textsc{MASPRM} (Table~\ref{tab:masprm-hitk}, App.~\ref{app:hitk-sensitivity}) and run a GSM8K \textsc{DyLAN}-inspired dynamic scheduler experiment (App.~\ref{app:dylan-dynamic}).

\vspace{-4pt}
\subsection{Overall experimental setup}
\vspace{-4pt}
\label{sec:exp-setup}

We evaluate on \textsc{GSM8K}~\cite{cobbe2021training} and \textsc{MATH}~\cite{hendrycksmath2021} for free-form math reasoning, and on \textsc{MMLU}~\cite{hendryckstest2021} and \textsc{LogiQA}~\cite{liu2020logiqa} (details in App.~\ref{app:datasets}). These tasks are judged by extracted final answers, with extraction and normalization described in App.~\ref{app:answer_extraction}.
We use fixed per-dataset MAS configurations from App.~\ref{app:graph} (roles, depth $D$, graph). Unless stated otherwise, agents are \texttt{Qwen2.5-1.5B-Instruct}. Learned scorers (\textsc{MASPRM} and ORM) use \texttt{Qwen2.5-1.5B/7B} as the base model.
We compare policy likelihood guidance (App.~\ref{app:policy-likelihood}) with learned scorers. \textsc{MASPRM} $V_\phi$ scores intermediate MAS states, whereas the outcome reward model (ORM) $\Gamma$ scores only terminal outputs. Since ORM is defined only on completed transcripts, we use it for terminal scoring or MCTS leaf evaluation. We exclude ORM from SBS because SBS must rank intermediate successors at each depth. Completing each successor before ranking would change the compute budget. Both are trained from MAS-MCTS rollouts labeled with terminal reward. \textsc{MASPRM} uses Bradley-Terry preference ranking (Sec.~\ref{sec:masprm-training}), while ORM uses terminal outcome supervision (App.~\ref{app:orm}). Additional training details appear in App.~\ref{app:exp-details}. App.~\ref{app:additional-experiments} defines the additional experiments and reports their main observations. Training time details appear in App.~\ref{app:rm-train-time}.
All datasets and pretrained model checkpoints used in this work are publicly released research artifacts, and we use them only for scientific training and evaluation under their respective licenses and terms of use.

For each dataset, we keep prompts and graphs fixed (App.~\ref{app:graph}) and vary only the inference algorithm and the scorer. Scorers are $\tilde\psi_{\text{pol}}$, $\Gamma$, or $V_\phi$ (Sec.~\ref{subsec:inference}). Single-Pass executes the MAS schedule once and returns the final extracted answer. Self-Consistency@$K$ (SC@$K$)~\cite{wang2023selfconsistency} runs $K{=}5$ independent MAS rollouts and selects the final answer by unweighted or scorer-weighted aggregation (App.~\ref{app:exp-details}).
SBS$(B_2,B_1)$ keeps a beam of width $B_1$, expands each beam state with $B_2$ sampled actions per step, and returns the highest scoring terminal candidate. MCTS($N_{\mathrm{sim}},C_{\max}$) runs UCT search and decodes by following the child with maximal backed up value from the initial state \(s_1(x)\) (Sec.~\ref{subsec:inference}, App.~\ref{app:algorithms}).

We report Hit@1 (exact-match after normalization), token cost (TOK ($\times 10^3$)), and agent calls (AC). Budget comparisons use matched within-method AC/TOK: variants inside a given SC@5, SBS, or MCTS block share the same agent model with comparable AC/TOK. We do not treat different search methods (e.g., SC@5 vs.\ MCTS) as matched AC/TOK comparisons. Reward model calls (RC) are deterministically recoverable from AC and inference hyperparameters; App.~\ref{app:budget} gives details.
For methods that output multiple ranked terminal candidates, we also report Hit@$k$ for $k \in \{1,3,5\}$ (Sec.~\ref{sec:exp-ablations}), i.e., the fraction of examples whose correct answer appears among the top-$k$ candidates.
The ranked candidates are complete terminal paths (full routed transcripts). For step-level scorers such as \textsc{MASPRM} or policy likelihood, the terminal candidate score is the mean of the step scores along the path. Terminal-only scorers score the completed transcript.
Unless stated, results use seed 42 and we average stochastic methods over three seeds (14, 24, 42).
Hardware details are in App.~\ref{app:hardware}.

\vspace{-4pt}
\subsection{\textsc{MASPRM} improves MAS within budget}
\vspace{-4pt}
\label{sec:exp-a}

We evaluate \textsc{MASPRM} across datasets and inference procedures (Single-Pass, SC@5, SBS(5,1), SBS(5,3), MCTS(10,3)). We compare three scoring signals: policy likelihood, an ORM, and \textsc{MASPRM} (with 1.5B vs.\ 7B scorer sizes where applicable).
Table~\ref{tab:main-results-masprm-qwen15b} reports Hit@1, TOK, and AC. Within each search-method block, the agent is fixed across rows (\texttt{Qwen2.5-1.5B-Instruct}), and scorer variants use comparable AC/TOK. Cross-method rows are context only.
\newcommand{\meanstd}[2]{%
  #1\,{\text{\fontsize{4pt}{4pt}\selectfont(\textpm\,#2)}}%
}

\begin{table*}[htbp]
  \centering
  \scriptsize
  \setlength{\tabcolsep}{3pt}
  \caption{Main results with \texttt{Qwen2.5-1.5B-Instruct} agents: Hit@1, TOK ($\times 10^3$), and AC (mean$\pm$std over 3 seeds). AC/TOK are comparable within each block. In the Scorer column, PL denotes policy likelihood, O denotes ORM, and M denotes \textsc{MASPRM}; the SC@5 row with no scorer (\texttt{--}) uses unweighted majority vote.}
  \label{tab:main-results-masprm-qwen15b}
  \vspace{-2pt}
  \resizebox{\textwidth}{!}{
  \begin{tabular}{ll
                  c >{\tiny}c >{\tiny}c    
                  c >{\tiny}c >{\tiny}c    
                  c >{\tiny}c >{\tiny}c    
                  c >{\tiny}c >{\tiny}c}   
    \toprule
    \multirow{2}{*}{Search} &
    \multirow{2}{*}{Scorer} &
    \multicolumn{3}{c}{\textsc{GSM8K}} &
    \multicolumn{3}{c}{\textsc{MATH}} &
    \multicolumn{3}{c}{\textsc{MMLU}} &
    \multicolumn{3}{c}{\textsc{LogiQA}} \\[1pt]
    \cmidrule(lr){3-5}
    \cmidrule(lr){6-8}
    \cmidrule(lr){9-11}
    \cmidrule(lr){12-14}
      & &
      Hit@1 & TOK & AC &
      Hit@1 & TOK & AC &
      Hit@1 & TOK & AC &
      Hit@1 & TOK & AC \\
    \midrule

    Single-Pass & -- &
      \meanstd{47.2}{1.9} & \meanstd{0.7}{0.0} & 4 &
      \meanstd{31.0}{1.5} & \meanstd{1.1}{0.1} & 4 &
      \meanstd{48.4}{0.6} & \meanstd{0.1}{0.0} & 3 &
      \meanstd{32.5}{1.2} & \meanstd{0.5}{0.0} & 3 \\

    \midrule
    SC@5 & -- &
      \meanstd{63.4}{1.7} & \meanstd{3.6}{0.0} & 20 &
      \meanstd{42.4}{0.9} & \meanstd{5.6}{0.1} & 20 &
      \meanstd{52.7}{0.6} & \meanstd{0.3}{0.1} & 15 &
      \meanstd{35.4}{1.9} & \meanstd{2.6}{0.1} & 15 \\

    SC@5 & PL &
      \meanstd{61.2}{1.3} & \meanstd{3.6}{0.0} & 20 &
      \meanstd{41.5}{0.9} & \meanstd{5.6}{0.1} & 20 &
      \meanstd{54.5}{1.9} & \meanstd{0.3}{0.1} & 15 &
      \meanstd{35.9}{0.8} & \meanstd{2.6}{0.1} & 15 \\

    SC@5 & O-1.5B &
      \meanstd{65.4}{1.1} & \meanstd{3.6}{0.0} & 20 &
      \meanstd{45.4}{1.3} & \meanstd{5.6}{0.1} & 20 &
      \meanstd{58.5}{0.7} & \meanstd{0.3}{0.1} & 15 &
      \meanstd{36.1}{1.2} & \meanstd{2.6}{0.1} & 15 \\

    \rowcolor{MASPRMRowBlue}
    SC@5 & M-1.5B &
      \meanstd{64.2}{0.9} & \meanstd{3.6}{0.0} & 20 &
      \meanstd{44.8}{1.0} & \meanstd{5.6}{0.1} & 20 &
      \meanstd{58.3}{0.2} & \meanstd{0.3}{0.1} & 15 &
      \meanstd{36.1}{0.7} & \meanstd{2.6}{0.1} & 15 \\

    SC@5 & O-7B &
      \meanstd{66.6}{0.9} & \meanstd{3.6}{0.0} & 20 &
      \meanstd{47.8}{1.0} & \meanstd{5.6}{0.1} & 20 &
      \meanstd{63.1}{1.4} & \meanstd{0.3}{0.1} & 15 &
      \meanstd{41.6}{0.8} & \meanstd{2.6}{0.1} & 15 \\

    \rowcolor{MASPRMRowBlue}
    SC@5 & M-7B &
      \textbf{\meanstd{67.8}{0.6}} & \meanstd{3.6}{0.0} & 20 &
      \textbf{\meanstd{49.2}{1.3}} & \meanstd{5.6}{0.1} & 20 &
      \textbf{\meanstd{67.2}{3.9}} & \meanstd{0.3}{0.1} & 15 &
      \textbf{\meanstd{43.0}{0.6}} & \meanstd{2.6}{0.1} & 15 \\

    \midrule
    SBS(5,1) & PL &
      \meanstd{55.0}{1.4} & \meanstd{3.6}{0.1} & 20 &
      \meanstd{41.0}{1.2} & \meanstd{5.5}{0.1} & 20 &
      \meanstd{55.0}{2.4} & \meanstd{0.3}{0.0} & 15 &
      \meanstd{35.9}{1.6} & \meanstd{2.6}{0.1} & 15 \\

    \rowcolor{MASPRMRowBlue}
    SBS(5,1) & M-1.5B &
      \meanstd{63.4}{0.9} & \meanstd{3.4}{0.1} & 20 &
      \meanstd{42.5}{1.6} & \meanstd{5.6}{0.1} & 20 &
      \meanstd{58.7}{1.6} & \meanstd{0.3}{0.0} & 15 &
      \meanstd{36.1}{0.7} & \meanstd{2.7}{0.1} & 15 \\

    \rowcolor{MASPRMRowBlue}
    SBS(5,1) & M-7B &
      \textbf{\meanstd{73.4}{0.7}} & \meanstd{3.2}{0.1} & 20 &
      \textbf{\meanstd{50.4}{0.9}} & \meanstd{5.5}{0.1} & 20 &
      \textbf{\meanstd{72.4}{1.7}} & \meanstd{0.2}{0.0} & 15 &
      \textbf{\meanstd{44.1}{0.9}} & \meanstd{2.6}{0.1} & 15 \\

    \midrule
    SBS(5,3) & PL &
      \meanstd{60.3}{1.2} & \meanstd{8.4}{0.1} & 50 &
      \meanstd{40.4}{1.0} & \meanstd{12.3}{0.2} & 50 &
      \meanstd{59.4}{1.8} & \meanstd{0.6}{0.1} & 35 &
      \meanstd{37.0}{0.7} & \meanstd{6.7}{0.0} & 35 \\

    \rowcolor{MASPRMRowBlue}
    SBS(5,3) & M-1.5B &
      \meanstd{69.1}{0.3} & \meanstd{8.4}{0.0} & 50 &
      \meanstd{41.2}{1.2} & \meanstd{12.3}{0.2} & 50 &
      \meanstd{60.4}{0.6} & \meanstd{0.8}{0.1} & 35 &
      \meanstd{37.3}{0.2} & \meanstd{6.6}{0.0} & 35 \\

    \rowcolor{MASPRMRowBlue}
    SBS(5,3) & M-7B &
      \textbf{\meanstd{77.5}{0.4}} & \meanstd{7.9}{0.0} & 50 &
      \textbf{\meanstd{51.3}{1.2}} & \meanstd{12.1}{0.2} & 50 &
      \textbf{\meanstd{74.8}{3.6}} & \meanstd{0.7}{0.1} & 35 &
      \textbf{\meanstd{45.0}{0.8}} & \meanstd{6.5}{0.0} & 35 \\

    \midrule
    MCTS(10,3) & PL &
      \meanstd{62.5}{0.7} & \meanstd{5.7}{0.1} & \meanstd{38.2}{5.4} &
      \meanstd{42.5}{1.1} & \meanstd{9.8}{0.3} & \meanstd{34.5}{3.2} &
      \meanstd{60.5}{0.8} & \meanstd{0.2}{0.0} & \meanstd{12.4}{1.0} &
      \meanstd{36.5}{1.1} & \meanstd{2.4}{0.1} & \meanstd{13.6}{2.1} \\

    MCTS(10,3) & O-1.5B &
      \meanstd{66.8}{0.6} & \meanstd{5.8}{0.1} & \meanstd{35.2}{3.2} &
      \meanstd{45.3}{1.1} & \meanstd{9.8}{0.3} & \meanstd{33.6}{4.1} &
      \meanstd{66.2}{0.7} & \meanstd{0.2}{0.0} & \meanstd{12.1}{1.3} &
      \meanstd{40.2}{0.9} & \meanstd{2.3}{0.1} & \meanstd{13.1}{2.6} \\

    \rowcolor{MASPRMRowBlue}
    MCTS(10,3) & M-1.5B &
      \meanstd{69.8}{0.5} & \meanstd{5.6}{0.1} & \meanstd{34.7}{2.4} &
      \meanstd{47.3}{0.9} & \meanstd{9.5}{0.3} & \meanstd{33.7}{3.5} &
      \meanstd{68.8}{0.6} & \meanstd{0.2}{0.0} & \meanstd{12.4}{1.9} &
      \meanstd{42.2}{0.8} & \meanstd{2.3}{0.1} & \meanstd{13.5}{1.7} \\

    MCTS(10,3) & O-7B &
      \meanstd{68.4}{0.8} & \meanstd{5.7}{0.1} & \meanstd{34.5}{2.8} &
      \meanstd{46.2}{0.9} & \meanstd{9.7}{0.3} & \meanstd{33.2}{3.3} &
      \meanstd{67.1}{0.6} & \meanstd{0.2}{0.0} & \meanstd{12.2}{1.1} &
      \meanstd{41.4}{0.8} & \meanstd{2.3}{0.1} & \meanstd{13.3}{2.2} \\

    \rowcolor{MASPRMRowBlue}
    MCTS(10,3) & M-7B &
      \textbf{\meanstd{82.9}{1.2}} & \meanstd{5.5}{0.1} & \meanstd{34.4}{1.7} &
      \textbf{\meanstd{51.8}{0.8}} & \meanstd{9.4}{0.3} & \meanstd{32.9}{2.7} &
      \textbf{\meanstd{75.2}{0.5}} & \meanstd{0.2}{0.0} & \meanstd{12.4}{0.6} &
      \textbf{\meanstd{45.5}{0.7}} & \meanstd{2.4}{0.1} & \meanstd{13.5}{2.4} \\

    \bottomrule
  \end{tabular}
  }
\end{table*}

\noindent Table~\ref{tab:main-results-masprm-qwen15b} shows different patterns by search method and scorer size. 

\textbf{Matched-size MCTS.} The scorer-vs-scorer comparison at matched scorer size under MCTS$(10,3)$ is consistently positive on all four datasets: \textsc{MASPRM} exceeds ORM by $+2.0$ to $+3.0$ points at 1.5B and $+4.1$ to $+14.5$ at 7B under comparable MCTS budget. 
\textbf{SC@5.} \textsc{MASPRM-1.5B} improves over policy likelihood on all four datasets, but ORM-1.5B is tied or ahead by 0.0 to 1.2 points. 
\textbf{SBS.} \textsc{MASPRM-7B} improves over policy likelihood on all four datasets for both SBS(5,1) and SBS(5,3), while \textsc{MASPRM-1.5B} gives smaller gains, especially on \textsc{MATH} and \textsc{LogiQA}; for SBS(5,1), these two gains are 1.5 and 0.2 points, and for SBS(5,3), they are 0.8 and 0.3 points. 
\textbf{Scorer scaling.} \textsc{MASPRM-7B} with fixed 1.5B agents improves over policy likelihood by an average of +13.4 under MCTS$(10,3)$, with comparable per-method averages of +13.4 for SBS(5,1), +12.9 for SBS(5,3), and +8.5 for SC@5. Figure~\ref{fig:gsm8k_tradeoff} visualizes the \textsc{GSM8K} Hit@1 versus token-cost tradeoff for these scorer and search choices. 
\textbf{Cost.} Although \textsc{MASPRM} adds reward-model calls, one scorer call is much cheaper than one agent generation on \textsc{GSM8K} (0.16s vs.\ 18.94s; App.~\ref{app:wall-clock}).

\textbf{Small-agent MAS vs.\ larger single-agent baselines.} As a contextual, non-compute-matched scale check (App.~\ref{app:single-agent-context}), MCTS(10,3)+\textsc{MASPRM} with 1.5B agents improves over the direct single-agent \texttt{Qwen2.5-1.5B-Instruct} baseline on all four tasks (82.9/51.8/75.2/45.5 vs. 41.9/17.5/34.3/25.8 for \textsc{GSM8K}/\textsc{MATH}/\textsc{MMLU}/\textsc{LogiQA}), but on \textsc{LogiQA} remains below \texttt{GPT-4o-mini} (50.1) and \texttt{Mistral~24B-Instruct} (56.4); see Table~\ref{tab:large-single-model-vs-mas} in App.~\ref{app:single-agent-context}.

\vspace{-4pt}
\subsection{Zero-shot OOD: Datasets \& Agents}
\vspace{-4pt}
\label{sec:exp-c}

We study two forms of zero-shot generalization.

\vspace{-4pt}
\subsubsection{Cross-dataset transfer}
\vspace{-4pt}
\label{sec:exp-cross-dataset}
For cross-dataset transfer, we train \textsc{MASPRM-7B} with a Bradley-Terry head on one dataset and evaluate it on a different related dataset, without reward-model retraining on the target dataset, under MCTS(10,3) with \texttt{Qwen2.5-1.5B-Instruct} as the agent. We focus on transfers within a domain, specifically math (\textsc{MATH} and \textsc{GSM8K}) and knowledge/reading (\textsc{MMLU} and \textsc{LogiQA}). Diagonal entries correspond to in-domain reward-model training. Results are reported in Table~\ref{tab:masprm-ood-with-logprob}.

\begin{table}[t]
  \centering
  \scriptsize
  \setlength{\tabcolsep}{6pt}
  \caption{Cross-dataset transfer under MCTS$(10,3)$: \textsc{MASPRM-7B} Hit@1 (\%) with train datasets as rows and test datasets as columns. Policy likelihood is the no learned reward model baseline.}
  \label{tab:masprm-ood-with-logprob}
  \vspace{-2pt}

  \begin{tabular}{lcccc}
    \toprule
    \multirow{2}{*}{Train dataset} & \multicolumn{4}{c}{Test dataset (Hit@1 \%)} \\
    \cmidrule(lr){2-5}
      & \textsc{GSM8K} & \textsc{MATH} & \textsc{MMLU} & \textsc{LogiQA} \\
    \midrule
    Policy Likelihood & 62.5 & 42.5 & 60.5 & 36.5 \\
    \midrule
    \textsc{GSM8K}  & \textbf{82.9} & 47.9 & --    & --    \\
    \textsc{MATH} & 73.6          & \textbf{51.8} & --    & --    \\
    \textsc{MMLU} & --            & --    & \textbf{75.2} & 41.8 \\
    \textsc{LogiQA} & --            & --    & 69.1 & \textbf{45.5} \\
    \bottomrule
  \end{tabular}
\end{table}

Training on one dataset and evaluating on another within the same domain yields cross-transfer performance that surpasses policy likelihood baselines, yet remains below in-domain training. Math transfer (between GSM8K and MATH) shows $-3.9$ to $-9.3$ point degradation vs in-domain but $+5.4$ to $+11.1$ point gain over policy likelihood (Table~\ref{tab:masprm-ood-with-logprob}). Knowledge/reading transfer (between MMLU and LogiQA) shows $-3.7$ to $-6.1$ point degradation vs in-domain but $+5.3$ to $+8.6$ point gain over policy likelihood. Averaged across transfers, cross-dataset within-domain \textsc{MASPRM} degrades by 5.8 points vs in-domain training but improves by +7.6 points vs policy likelihood.

\vspace{-4pt}
\subsubsection{Cross-agent transfer}
\vspace{-4pt}
\label{sec:exp-cross-agent}
For cross-agent transfer, we train reward models on \texttt{Qwen2.5-1.5B-Instruct} rollouts and swap the generator agents to \texttt{Llama3.2~3B-Instruct} at inference. The reward models remain fixed, without retraining (Sec.~\ref{sec:exp-setup}). We repeat the search/scoring setups from Sec.~\ref{sec:exp-a} and report results in Table~\ref{tab:main-results-masprm-cross-llm} in App.~\ref{app:cross-llm}.
\textsc{MASPRM-7B} trained on Qwen2.5-1.5B rollouts transfers to \texttt{Llama3.2~3B-Instruct} agents without retraining, despite differences in tokenizer, pretraining corpus, and base architecture between the two model families. Under MCTS(10,3), it improves over policy likelihood by +8.7 to +13.0 points across datasets and over ORM-7B by +4.0 to +8.0 points (Table~\ref{tab:main-results-masprm-cross-llm} in App.~\ref{app:cross-llm}).

\vspace{-4pt}
\subsection{Ablations}
\vspace{-4pt}
\label{sec:exp-ablations}
\textbf{Reward head.} Table~\ref{tab:masprm-head-ablation} in App.~\ref{app:head-ablation} compares reward head choices under MCTS(10,3) with \texttt{Qwen2.5-1.5B-Instruct} agents. BT is the best \textsc{MASPRM} head on all four datasets for both scorer sizes, and BCE is the best or tied ORM head. With size-matched 1.5B models and optimal heads, \textsc{MASPRM} is 2.0 to 3.0 points above ORM across datasets under the same MCTS(10,3) setting. 

\textbf{Scorer observability.} Table~\ref{tab:masprm-local-observability} in App.~\ref{app:local-observability} tests scorer observability in the same setting. The full routed transcript beats the local view variant on every dataset by 2.2 to 5.5 points: 82.9 vs.\ 77.4 on \textsc{GSM8K}, 51.8 vs.\ 49.4 on \textsc{MATH}, 75.2 vs.\ 71.7 on \textsc{MMLU}, and 45.5 vs.\ 43.3 on \textsc{LogiQA} in the same MCTS(10,3) setting.

\textbf{Dynamic-scheduler feasibility.} As an exploratory check, Table~\ref{tab:gsm8k-dylan-dynamic} in App.~\ref{app:dylan-dynamic} evaluates a \textsc{GSM8K} \textsc{DyLAN}-style scheduler. The dynamic scheduler has higher mean Hit@1 than scheduled MAS$_1$ in Single-Pass (51.2 vs.\ 47.2) and MCTS(10,3) (84.6 vs.\ 82.9), but MCTS cost rises substantially, from 5.5 to 11.9 TOK ($\times 10^3$) and from 34.4 to 93.8 AC. We therefore treat this result as preliminary evidence that \textsc{MASPRM} can be applied beyond the main predefined MAS configurations, not as a broad claim about dynamic scheduling.

\textbf{Ranking stability and graph sensitivity.} Table~\ref{tab:masprm-hitk} in App.~\ref{app:hitk-sensitivity} and Table~\ref{tab:gsm8k-graph-sensitivity} in App.~\ref{app:graph-sensitivity} test ranking stability and graph sensitivity. \textsc{MASPRM-7B} reduces the average Hit@1 to Hit@5 gap from 16.1 to 5.8 under MCTS and from 19.0 to 10.5 under SC@5, indicating better top-ranked selection. Across three \textsc{GSM8K} graphs, guided search ranges are 1.8 to 2.2 points, compared with a 9.0 point Single-Pass range, indicating reduced sensitivity to graph choice.

\vspace{-6pt}
\section{Discussion}
\vspace{-6pt}

\textbf{RQ1.} The main supported claim is that \textsc{MASPRM} improves inference-time search in the predefined MAS configurations studied here for verifiable terminal-answer tasks (Sec.~\ref{sec:exp-a}). The evidence is strongest for stepwise search, especially MCTS, where the scorer is queried on intermediate routed transcripts and can affect which branches are expanded or pruned (Sec.~\ref{sec:exp-a}, Table~\ref{tab:main-results-masprm-qwen15b}). When search returns a ranked list, \textsc{MASPRM} also narrows Hit@1 to Hit@5 gaps (Sec.~\ref{sec:exp-ablations} and Table~\ref{tab:masprm-hitk} in App.~\ref{app:hitk-sensitivity}), suggesting better ranking of routed transcripts. The local-observability ablation suggests that the multi-agent state representation is part of the gain rather than only a generic answer-scoring signal (Sec.~\ref{sec:exp-ablations} and Table~\ref{tab:masprm-local-observability} in App.~\ref{app:local-observability}).

\textbf{RQ2.} \textsc{MASPRM} is not uniformly better than ORM across all inference procedures (Sec.~\ref{sec:exp-a}). The size-matched MCTS rows are the cleaner comparison to ORM and are consistently positive across all four datasets: $+2.0$ to $+3.0$ points at 1.5B and $+4.1$ to $+14.5$ at 7B; the 7B-vs-policy-likelihood gains additionally reflect scorer scaling with fixed 1.5B agents (Sec.~\ref{sec:exp-a}). At 1.5B, SC@5 is mixed: \textsc{MASPRM} improves over policy likelihood but does not beat ORM. This pattern is consistent with the role of each scorer. SC@5 primarily selects among final answers, which is closer to the outcome-level task ORM is designed for, while \textsc{MASPRM} is designed for stepwise guidance. Thus, the mixed SC@5 result does not contradict the stronger MCTS gains because the two procedures use the scorers in different ways. Together, the MCTS results and head ablation support the narrower claim that process guidance is most useful when the algorithm repeatedly makes intermediate search decisions (Sec.~\ref{sec:exp-a} and Sec.~\ref{sec:exp-ablations}). They do not show that \textsc{MASPRM} universally replaces ORM for final-answer selection. Head ablations also suggest that preference ranking heads are a particularly good fit for search because they mirror the pairwise comparisons performed during selection (Sec.~\ref{sec:exp-ablations}, Table~\ref{tab:masprm-head-ablation} in App.~\ref{app:head-ablation}).

\textbf{RQ3.} The learned process signal generalizes beyond a single rollout distribution: \textsc{MASPRM} transfers nontrivially across related datasets and one different generator model without retraining, while in-domain reward-model training remains better (Sec.~\ref{sec:exp-c}, Table~\ref{tab:masprm-ood-with-logprob}, and Table~\ref{tab:main-results-masprm-cross-llm} in App.~\ref{app:cross-llm}). The graph-sensitivity ablation gives the same direction across three GSM8K communication graphs (Sec.~\ref{sec:exp-ablations}, Table~\ref{tab:gsm8k-graph-sensitivity} in App.~\ref{app:graph-sensitivity}), so the effect is not limited to one tested graph. It suggests that guided search can reduce, though not eliminate, sensitivity to the particular MAS graph among the tested GSM8K configurations.

Finally, App.~\ref{app:single-agent-context} shows guided small-agent MAS is strong on GSM8K/MMLU and roughly competitive on MATH (Table~\ref{tab:large-single-model-vs-mas}). On \textsc{LogiQA}, the strongest single-agent baselines remain ahead, indicating that inference-time MAS scaling complements but does not yet substitute for large-model scaling on harder reasoning tasks. Together, these results suggest that inference-time multi-agent search becomes substantially more reliable when guided by a learned process value model trained from terminal rewards alone, offering dense feedback without human step-level annotations.

\vspace{-6pt}
\section{Conclusion}
\vspace{-6pt}

We presented \textsc{MASPRM}, a process reward model for multi-agent systems on verifiable terminal-answer tasks. It scores routed transcript prefixes and guides SBS and MCTS. Across four benchmarks, \textsc{MASPRM} improves predefined MAS search over policy likelihood under comparable within-method budgets. Relative to ORM, the comparison depends on the inference method: SC@5 at 1.5B is mixed, while stepwise search that uses intermediate decisions, especially MCTS, benefits more from \textsc{MASPRM}, exceeding size-matched ORM by $+2.0$ to $+3.0$ points at 1.5B and $+4.1$ to $+14.5$ at 7B across the four benchmarks. The 7B-vs-policy-likelihood gains additionally reflect scorer scaling. \textsc{MASPRM} also transfers across related datasets and agent LLMs without retraining, suggesting that learned process-level value signals can be a useful component for MAS search.

\vspace{-6pt}
\section{Limitations}
\vspace{-6pt}
We scope our study to predefined, role-pipelined MAS configurations on verifiable terminal-answer tasks, where a learned process value model has a well-defined supervision signal; extending to open-ended generation, where terminal correctness is undefined, is left to future work. Extending \textsc{MASPRM} to mixed-initiative or dynamically structured MAS, where routing and termination depend on intermediate outputs, is a natural next step; App.~\ref{app:dylan-dynamic} reports one preliminary instance. The default scorer conditions on the full routed transcript, which fits centralized inference; our local-observability ablation (Sec.~\ref{sec:exp-ablations}) shows that the full-transcript scorer outperforms a local-view variant by 2.2 to 5.5 Hit@1 points, motivating future work on decentralized scorers that close this gap. Inference-time scoring cost scales with transcript length, but remains substantially below per-agent generation cost (App.~\ref{app:wall-clock}, e.g., 0.16s scorer call vs.\ 18.94s agent call on \textsc{GSM8K}).

\vspace{-6pt}
\section{Ethical Considerations}
\vspace{-6pt}
\textsc{MASPRM} is a research method for guiding inference-time search in predefined MAS on verifiable terminal-answer benchmarks, not a guarantee of correctness or safety. Because it ranks intermediate routed transcripts, it could encourage trust in a flawed trajectory without task-specific validation, uncertainty reporting, and human oversight, especially in high-stakes use. It also adds inference-time computation through extra agent generations and scorer calls, so use should weigh accuracy gains against latency, energy use, and environmental cost. Our experiments use public datasets and pretrained models under their licenses, collect no new human subject data, and do not redistribute third-party datasets, weights, or API outputs. Open-ended, tool-use, or user-facing agents may add privacy, security, and misuse risks beyond those studied here, such as sensitive routed messages and adversarial manipulation of the search controller.

\bibliography{custom}

\newpage
\MASPRMrestoreSpacing

\appendix

\section{Illustrative Examples}
\label{app:examples}

\subsection{Example MAS and schedule}
\begin{figure}[!ht]
  \centering
  \includegraphics[width=.72\linewidth]{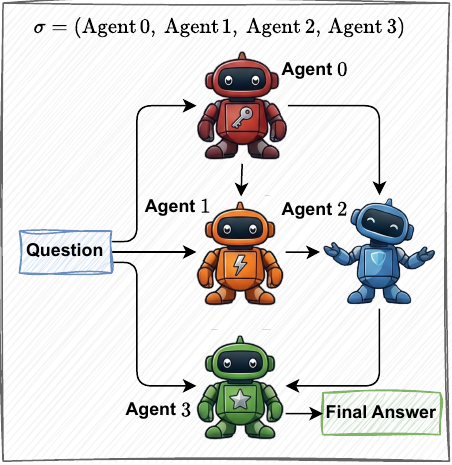}
  \caption{Illustrative four-agent MAS with a fixed schedule. The schedule orders one turn per agent, and directed edges define allowed routing from the question to the final answer.}
  \label{fig:mas}
\end{figure}

Fig.~\ref{fig:mas} shows an illustrative four-agent MAS. Its schedule $\sigma$ orders Agent 0, Agent 1, Agent 2, and Agent 3. The directed edges show allowed message routing from the question node through the agents to the final answer. This example does not exactly match all MAS variants used in experiments. See App.~\ref{app:graph} for the full set of MAS configurations used in the results and ablations.

\subsection{Stylized transcript for an arithmetic word problem}
\begin{figure*}[!ht]
  \centering
  \includegraphics[width=\textwidth]{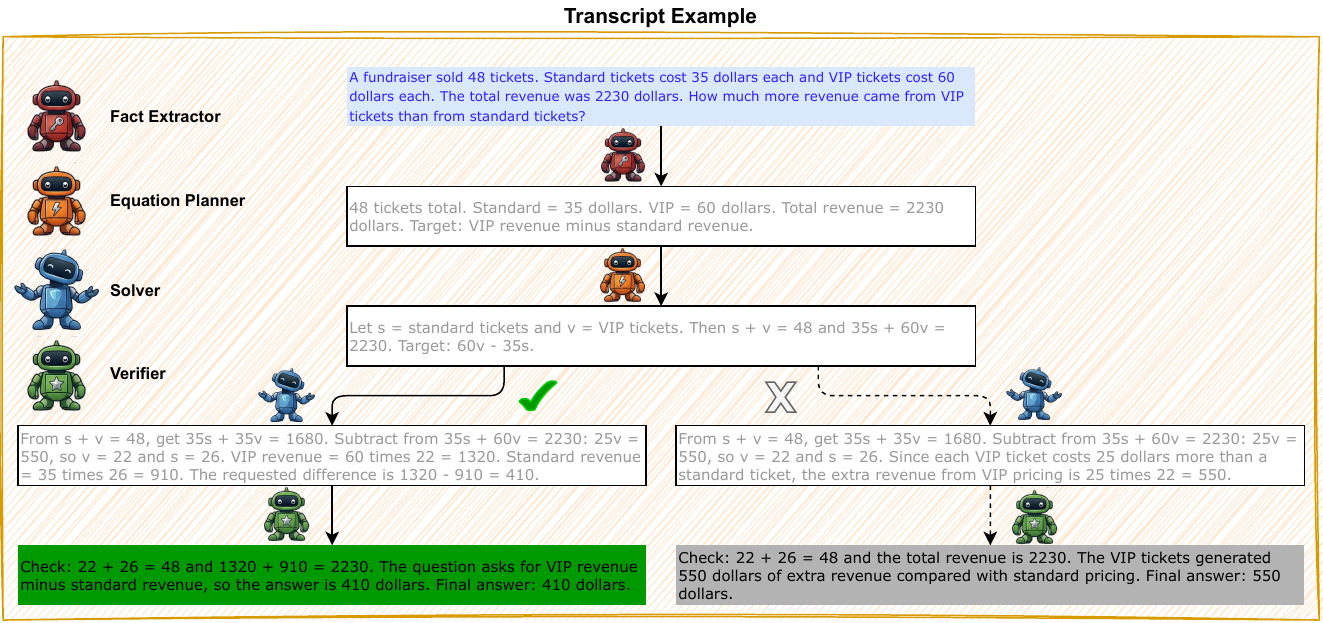}
  \caption{Example transcript trajectories for a ticket-revenue word problem. A four-agent MAS extracts facts, plans equations, solves the system, and verifies the final answer. The green trajectory keeps the requested quantity as VIP revenue minus standard revenue and reaches $1320{-}910=410$. The gray trajectory finds the same ticket counts but switches to a different quantity, extra revenue from VIP pricing, and returns $25{\times}22=550$.}
  \label{fig:transcript_example}
\end{figure*}

We visualize (Fig.~\ref{fig:transcript_example}) one rollout on a ticket-revenue problem: "A fundraiser sold 48 tickets. Standard tickets cost \$35 each and VIP tickets cost \$60 each. The total revenue was \$2230. How much more revenue came from VIP tickets than from standard tickets?"
The available outputs are from the \emph{Equation Planner} and \emph{Solver}. The green trajectory keeps the requested target as VIP revenue minus standard revenue and returns $410$. The gray trajectory solves the same system but answers a different quantity, the extra revenue from VIP pricing, and returns $550$.

\subsection{MCTS backpropagation example}
\begin{figure}[!ht]
  \centering
  \includegraphics[width=.82\linewidth]{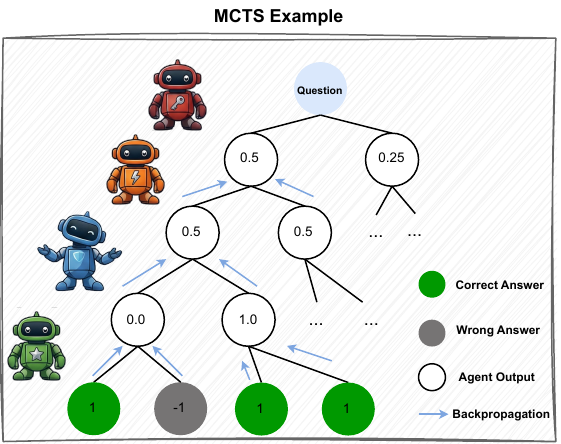}
  \caption{Example of MCTS backpropagation in MAS-MCTS data generation. Terminal leaves are labeled by the final answer reward, with green leaves for correct answers ($R=+1$) and gray leaves for wrong answers ($R=-1$). Blue arrows show how rewards are backed up along visited paths to update empirical values for earlier agent outputs. These values provide the sibling comparisons used to train \textsc{MASPRM}.}
  \label{fig:mcts}
\end{figure}

Fig.~\ref{fig:mcts} illustrates MCTS backpropagation on an unrolled MAS tree. Terminal leaves are scored by the final answer reward, with green for correct answers ($+1$) and gray for wrong answers ($-1$). The blue arrows show reward backup from leaves to earlier agent outputs, producing empirical values such as $0.5$, $1.0$, and $0.25$. These backed-up values define the process targets used to compare sibling successor states in \textsc{MASPRM} training.

\section{Additional Experimental Details}
\label{app:exp-details}

\subsection{Benchmarks and Dataset Splits}
\label{app:datasets}

We use the four benchmarks from Sec.~\ref{sec:exp-setup}.
All datasets are loaded with the HuggingFace \texttt{datasets} library. When forming subsets, we shuffle with a fixed random seed (42).
For each dataset, we define a training pool $\mathcal{D}_{\mathrm{train}}$ and a disjoint held out evaluation set $\mathcal{D}_{\mathrm{eval}}$ used for reporting.
We then create a development set $\mathcal{D}_{\mathrm{dev}}$ (for hyperparameter selection) by holding out 5\% of the (shuffled) training pool:
\begin{equation}
|\mathcal{D}_{\mathrm{dev}}| = \left\lfloor 0.05\,|\mathcal{D}_{\mathrm{train}}|\right\rfloor,
\,
\mathcal{D}_{\mathrm{train}} \leftarrow \mathcal{D}_{\mathrm{train}} \setminus \mathcal{D}_{\mathrm{dev}}.
\end{equation}

For MATH (Competition-Math), we use a held out evaluation subset $\mathcal{D}_{\mathrm{eval}}$ with $|\mathcal{D}_{\mathrm{eval}}|{=}2000$ examples and use the remainder as the training pool, holding out 5\% of $\mathcal{D}_{\mathrm{train}}$ as a development set.

For GSM8K and LogiQA, we use the \texttt{train} split as $\mathcal{D}_{\mathrm{train}}$ and the \texttt{test} split as $\mathcal{D}_{\mathrm{eval}}$, and we hold out 5\% of $\mathcal{D}_{\mathrm{train}}$ as a development set (we do not use any provided validation split).

For MMLU, we use the larger provided partition as $\mathcal{D}_{\mathrm{train}}$ and the other partition as $\mathcal{D}_{\mathrm{eval}}$, and we hold out 5\% of $\mathcal{D}_{\mathrm{train}}$ as a development set.

Finally, we optionally cap the number of used examples to $n$ by taking the first $n$ examples after shuffling. Table~\ref{tab:dataset-splits} in App.~\ref{app:datasets} reports the exact split sizes after applying this protocol.

\begin{table}[t]
\centering
\small
\setlength{\tabcolsep}{6pt}
\begin{tabular}{lrrr}
\toprule
Dataset & Train & Dev & Eval \\
\midrule
MATH (Competition-Math) & 9{,}975 & 525 & 2{,}000 \\
GSM8K                  & 7{,}100 & 373 & 1{,}319 \\
MMLU  & 13{,}340 & 702 & 1{,}531 \\
LogiQA                 & 7{,}008 & 368 & 651 \\
\bottomrule
\end{tabular}
\caption{Dataset split sizes (Train/Dev/Eval).}
\label{tab:dataset-splits}
\end{table}

\subsection{Answer extraction and normalization}\label{app:answer_extraction}
Final answers are extracted from the terminal agent output (e.g., \texttt{Verifier}, \texttt{Judge}, \texttt{FinalJudge}, \texttt{Query\_Checker}, or the single-agent baseline), with multiple fallback patterns to accommodate the different MAS prompt templates used in App.~\ref{app:graph}.

We first attempt to extract a final answer span using the following regex over common final answer templates:
\[
\small
\texttt{(?i)\textbackslash b(?:Final Answer|Answer)\textbackslash s*:?\textbackslash s*(.+)}.
\]
If a match is found, we use the captured group as the candidate answer.

If no explicit final answer tag is present, we fall back to:
(i) for free-form datasets (\textsc{GSM8K}, \textsc{MATH}): strip trailing units from the final message (e.g., \texttt{cm}, \texttt{degrees}, or \texttt{\textbackslash text\{...\}}) and then take the last remaining number-like token.
(ii) for multiple-choice datasets (\textsc{MMLU}, \textsc{LogiQA}): the first option label in $\{A,B,C,D\}$ (case insensitive) that appears as a standalone token in the final message.

For free-form datasets, we parse the captured span as a number when possible and apply light normalization (lowercase, strip commas/spaces, and remove trailing unit markers). We accept numeric answers within tolerance $\delta=10^{-3}$. For multiple-choice datasets, we map the extracted label to the canonical answer. Non-parsable outputs or timeouts are scored incorrect.

\subsection{\textsc{MASPRM} training data generation}
\label{app:masprm-data}
For training data generation, we run MAS-MCTS(40,3) on training problems (i.e., $N_{\mathrm{sim}}{=}40$ simulations and expansion cap $C_{\max}{=}3$ children per expanded state) with $c_{\textsc{uct}}{=}4.0$.
Leaf values are set to the ground-truth terminal reward, $v_{\mathrm{leaf}}(s_T)=R(s_T)\in\{-1,+1\}$ (no virtual visits during training).

\paragraph{Process targets.}
From the resulting search trees, we construct process targets at the post action state level.
For each visited edge $(s_t,a)$ with successor $s'=\mathrm{next}(s_t,a)$, we assign
$y_{\mathrm{proc}}(s') \triangleq \widehat Q(s_t,a)$ where $\widehat Q(s_t,a)=W(s_t,a)/N(s_t,a)$ is the backed up Monte Carlo estimate. Targets are clipped to $[-1,1]$.

\paragraph{Preference pairs (BT head).}
Eq.~\eqref{eq:d_pref} in the main text defines $\mathcal{D}_{\mathrm{pref}}$ as all ordered sibling pairs with $y_{\mathrm{proc}}(s^+)>y_{\mathrm{proc}}(s^-)$. In practice, materializing all such pairs can be quadratic in the number of expanded children and can introduce many near-duplicate comparisons (e.g., formatting variants of the same action). We therefore mine a small, high-signal subset of pairs from each MAS-MCTS tree and use it as the instantiated $\mathcal{D}_{\mathrm{pref}}$ in our experiments.
We fix the path length to $D{+}1$ nodes (the initial state plus $D$ agent actions) and, for each prefix node at depth $d$ ($0 \le d < D$), we compare next step candidate children that share the same prefix state. Concretely, at each such prefix we:
(i) deduplicate children by normalized action text,
(ii) rank unique candidates by their backed up value $y_{\mathrm{proc}}(s')$,
(iii) take up to $K_{+}{=}4$ top-valued and $K_{-}{=}4$ bottom-valued candidates (disjoint),
and (iv) emit pairs $(s^+,s^-)$ between the two sets when $y_{\mathrm{proc}}(s^+)>y_{\mathrm{proc}}(s^-)$ (ties ignored).
For the final step ($d{=}D{-}1$), we restrict candidates to terminal children.
We cap the total number of mined preference pairs per tree to $8$.

\paragraph{Pointwise samples (MSE head).}
For the MSE head ablations, we extract fixed length transcripts of $D{+}1$ nodes from the initial state to terminal states and supervise each step with the node value (clipped to $[-1,1]$). We keep only terminal leaves with reward $\pm 1$, and remove exact duplicate samples.

\subsection{\textsc{MASPRM} model and optimization}
Backbone: \texttt{Qwen2.5-1.5B} or \texttt{Qwen2.5-7B}, depending on scorer size, with a scalar head that maps each state to a single score in $[-1,1]$ (via a $\tanh$ output). For the BT variant, this scalar is the Bradley-Terry score used for pairwise ranking, and we use its raw value as the inference score. For the MSE variant, we regress the scalar to $y_{\mathrm{proc}}(s')$. We use QLoRA 4-bit with rank 256 and dropout 0.05. Batch size is $8$ (effective with grad accumulation $16$), LR $1{\times}10^{-5}$, epochs $5$, optimizer \texttt{ADAMW}, and weight decay $0.01$. Evaluation metrics on held out preference pairs include average margin, pairwise accuracy, AUC, and BT loss.

\subsection{Search configurations}
When a full rollout must be ranked, each candidate is the complete path from $s_1(x)$ to a terminal state. With \textsc{MASPRM}, the path score is the mean value after each action, averaged across turns. With policy likelihood, the path score is the mean policy score over the generated actions. ORMs score only the completed transcript.

SC@$K$: We run $K{=}5$ independent MAS rollouts that produce answers $\hat y^{(i)}$ and scalar path scores $w^{(i)}$. When no scorer is used, SC@5 uses unweighted majority vote. Otherwise, we sigmoid normalize $\{w^{(i)}\}_{i=1}^K$ to $\tilde w^{(i)}\in[0,1]$, compute a total score for each distinct answer
\(
S(\hat y)=\sum_{i:\hat y^{(i)}=\hat y} \tilde w^{(i)}
\),
and return the answer with the largest $S(\hat y)$.

SBS$(B_2,B_1)$: At each depth, we sample $B_2$ candidates per beam state and keep the top $B_1$ globally (we report SBS(5,1) and SBS(5,3)). Temperature $=0.7$, top-$p=0.95$. Tie-breaking uses the higher \textsc{MASPRM} score (if applicable), otherwise the higher final answer policy likelihood.

MCTS$(N_{\mathrm{sim}},C_{\max})$: Selection uses UCT (Eq.~\eqref{eq:uct}). In our main experiments we use MCTS(10,3) ($N_{\mathrm{sim}}{=}10$, $C_{\max}{=}3$). Expansion samples up to $C_{\max}$ children by top-$p$ sampling from the current agent. Leaf evaluation uses $V_\phi$. Rollout depth is capped at the scheduled depth $D$ of the active MAS configuration (App.~\ref{app:graph}), so that the final scheduled agent produces the answer. On dead ends, we back up the current \textsc{MASPRM} leaf value. No policy priors are added to UCT unless otherwise stated.
For the \textsc{DyLAN}-style dynamic experiment (App.~\ref{app:dylan-dynamic}), the same MCTS code follows the enabled conditional edges and stops at the active terminal output; all other reported experiments use the fixed depth $D$ above.

\subsection{Compute budgets and accounting}
\label{app:budget}
Within each search-method block, we use matched/comparable agent-call and token budgets. An agent call is one LLM generation by one agent at one depth. TOK counts total generated tokens aggregated over all agent calls. RC counts reward model forward passes used to score candidate successor states during search.

\paragraph{Recovering RC from AC.}
In our reported setups, RC is deterministically recoverable from AC and the inference hyperparameters (and the scheduled depth $D$, App.~\ref{app:graph}), so we omit RC from the main tables. Policy likelihood guidance uses no learned scorer, so $\mathrm{RC}=0$. When a scorer is applied to each newly generated successor state (SBS, MCTS with \textsc{MASPRM}, and SC@$K$ with \textsc{MASPRM}), we score each successor once, giving $\mathrm{RC}=\mathrm{AC}$. For ORMs used only to score terminal candidates, RC equals the number of terminal evaluations: SC@$K$ uses $\mathrm{RC}=K=\mathrm{AC}/D$, and MCTS$(N_{\mathrm{sim}},C_{\max})$ with an ORM at simulation leaves uses $\mathrm{RC}=N_{\mathrm{sim}}$ (one ORM call per simulation).

Let $D$ denote the scheduled depth (number of agent turns) in a single MAS pass. $D$ depends on the MAS configuration (App.~\ref{app:graph}). Single-Pass uses $\mathrm{AC}=D$, and SC@$K$ uses $\mathrm{AC}=K\,D$. For SBS with global beam $B_1$ and samples per state $B_2$, the agent call budget is
\(
\mathrm{AC}=B_2 + (D{-}1)\,B_1\,B_2
\).
For MCTS, $\mathrm{AC}$ is not deterministic due to stochasticity. We report the empirical mean $\pm$ std and tune hyperparameters to obtain comparable average AC and TOK across scorers.

\paragraph{Total compute reported for the checklist.}
Data generation uses MAS-MCTS(40,3), so a conservative upper bound is
$40C_{\max}D$ agent calls per training instance. For the four main
in-domain datasets, this is at most 15.5M agent calls total: 3.4M for
\textsc{GSM8K}, 4.8M for \textsc{MATH}, 4.8M for \textsc{MMLU}, and
2.5M for \textsc{LogiQA}; the separate \textsc{GSM8K}
\textsc{DyLAN}-style run adds at most 6.8M. Evaluation compute is obtained
by multiplying the per-example AC/TOK values in the result tables by the
evaluation-set sizes in Table~\ref{tab:dataset-splits}. Scorer fine-tuning
uses at most 384/576 V100-GPU hours per 1.5B/7B scorer
(App.~\ref{app:rm-train-time}); across the 37 scorer fine-tunes used for the main results and ablations, this gives an upper bound of 18.2k V100-GPU hours for scorer training.

\subsection{Hardware}
\label{app:hardware}
Experiments run on $8\times$ NVIDIA Tesla V100-PCIE-32GB (32\,GB VRAM, driver 535.54.03, CUDA 12.2), an Intel Xeon Gold 6140 CPU @ 2.30\,GHz, and 754\,GB RAM.

\subsection{MASPRM training time}
\label{app:rm-train-time}
As a minimal reference point, GSM8K has 7{,}100 training instances (Table~\ref{tab:dataset-splits} in App.~\ref{app:datasets}). With our cap of at most 8 preference pairs mined per MAS-MCTS tree (App.~\ref{app:masprm-data}), this yields at most 56{,}800 preference pairs for \textsc{MASPRM} fine-tuning. With 5 epochs and effective batch size 128 (batch size 8 with grad accumulation 16), this corresponds to $\approx 2.2{\times}10^3$ optimizer steps. On our V100 setup, fine-tuning a single scorer typically takes at most 48 hours for the 1.5B model and at most 72 hours for 7B. Other datasets are similar and scale roughly with the corresponding pair counts (56k--107k from Table~\ref{tab:dataset-splits} in App.~\ref{app:datasets}).

\subsection{Policy likelihood guidance}
\label{app:policy-likelihood}
To compare against policy likelihood guidance, we score each candidate action $a$ by its length-normalized log-likelihood under the acting agent's policy~\cite{yang2018breaking,koehn2017six}:
\begin{equation}
\label{eq:psi-pol}
\psi_{\text{pol}}(a\mid s_t) \;=\; \frac{1}{L}\sum_{\ell=1}^{L} \log \pi^{(i_t)}\!\big(z_\ell \mid z_{<\ell}, x, H_{t-1}^{(i_t)}\big),
\end{equation}
where $z_{1:L}$ are the $L$ generated tokens of candidate action $a$.
Because $\psi_{\text{pol}}$ has a different scale than learned value heads, we map it to $[0,1]$ with a sigmoid:
$\tilde\psi_{\text{pol}}(a\mid s_t)=\mathrm{sigmoid}(\psi_{\text{pol}}(a\mid s_t))$ where $\mathrm{sigmoid}(z)=1/(1+\exp(-z))$.
In MCTS, we use $\tilde\psi_{\text{pol}}$ for virtual-visit initialization $f_{\textsc{init}}(s_t,a)$ and for frontier leaf evaluation $v_{\mathrm{leaf}}(s_t)=\max_{a\in C_t}\tilde\psi_{\text{pol}}(a\mid s_t)$. In SBS, we rank successors by $\tilde\psi_{\text{pol}}$.

\subsection{\texorpdfstring{ORM $\Gamma$}{ORM Gamma}}
\label{app:orm}
We compare against a standard ORM that assigns a scalar score to a completed transcript. Concretely, an ORM $\Gamma$ maps a terminal state $s_T=(x,H_{T-1},\text{sink})$ (i.e., the full routed transcript, the same input \textsc{MASPRM} receives at the final step) to a scalar score and is only defined on terminal states (it is not used on intermediate states).

We train ORMs from the same MAS-MCTS rollouts used for \textsc{MASPRM}, but using only terminal outcomes for supervision. From each terminal state $s_T$, we compute the terminal reward $r=R(s_T)\in\{-1,+1\}$ (via the extracted final answer $\hat y(s_T)$) to form:
\small
\(
\mathcal{D}_{\mathrm{out}}=\big\{\,(s_T,\,r)\;:\; s_T \text{ terminal},\; r=R(s_T)\,\big\}.
\)
\normalsize
We then fit the ORM $\Gamma_{\theta_\Gamma}$ by minimizing
\[
\min_{\theta_\Gamma}\;\frac{1}{|\mathcal{D}_{\mathrm{out}}|}\sum_{(s_T,r)\in\mathcal{D}_{\mathrm{out}}}
\ell\!\big(\Gamma_{\theta_\Gamma}(s_T),\,r\big),
\]
using either a BCE classification head (treating correctness as a binary label) or an MSE regression head (regressing to $r\in\{-1,+1\}$). Training hyperparameters and optimizer mirror the \textsc{MASPRM} training. Loss/head variants (BCE vs.\ MSE) are detailed in App.~\ref{app:head-ablation}.

\subsection{Artifact licenses and intended use}
\label{app:artifact-licenses}
We use external artifacts only for scientific training and evaluation.
Datasets: \textsc{GSM8K} (MIT), \textsc{MATH} (MIT), \textsc{MMLU}
(MIT), and \textsc{LogiQA} (public research release; no explicit dataset
license found in the original repository or Hugging Face loader). Models:
\texttt{Qwen2.5-1.5B/7B} (Apache-2.0), \texttt{Llama3.1/3.2}
checkpoints (Llama Community License), Mistral 24B-Instruct (Apache-2.0),
and \texttt{GPT-4o-mini} through the OpenAI API under provider terms. We do
not redistribute third-party datasets, third-party weights, or API outputs.
Released code will use MIT; any released adapters/checkpoints will include
model cards preserving upstream model and dataset notices.

\section{Additional Experiments}
\label{app:additional-experiments}

This section gives the full setup for additional experiments referenced in the main text and states the main observation for each one.

\subsection{Contextual single-agent comparison}
\label{app:single-agent-context}

We include a scale-context comparison between direct single-agent baselines and guided MAS search.
A single-agent baseline means one instruction model answers the problem directly without MAS communication, multi-agent search, or a learned scorer.
The guided MAS rows use \texttt{Qwen2.5-1.5B-Instruct} generator agents and \textsc{MASPRM-7B}; the reported MCTS(10,3) and SBS(5,3) results are the same predefined MAS configurations used in the main results.
This comparison is not matched by AC/TOK and should not be read as evidence against optimized single-agent PRM+search systems or as a compute-matched comparison with larger single-agent models.

\begin{table}[!htbp]
  \centering
  \scriptsize
  \setlength{\tabcolsep}{6pt}
  \caption{Contextual Hit@1 (\%) comparison: direct single-agent baselines versus guided MAS with \texttt{Qwen2.5-1.5B-Instruct} agents and \textsc{MASPRM-7B}.}
  \label{tab:large-single-model-vs-mas}
  \vspace{-2pt}
  \resizebox{\columnwidth}{!}{
  \begin{tabular}{lcccc}
    \toprule
    Method & \textsc{GSM8K} & \textsc{MATH} & \textsc{MMLU} & \textsc{LogiQA} \\
    \midrule

    \multicolumn{5}{l}{\textbf{Single-Agent Baselines}} \\
    \midrule
    \texttt{Qwen2.5-1.5B-Instruct} & 41.9 & 17.5 & 34.3 & 25.8 \\
    \texttt{Llama3.2~3B-Instruct} & 54.8 & 24.4 & 41.2 & 35.3 \\
    \texttt{Llama3.1~8B-Instruct}    & 36.2 & 26.7 & 50.5 & 42.2 \\
    \texttt{GPT-4o-mini}  & 62.0 & 51.0 & 59.0 & 50.1 \\
    \texttt{Mistral~24B-Instruct}    & 80.4 & 51.9 & 60.6 & 56.4 \\
    \midrule
    \multicolumn{5}{l}{\textbf{Guided MAS (\texttt{Qwen2.5-1.5B-Instruct})}} \\
    \midrule
    MCTS(10,3)+\textsc{MASPRM} & \textbf{82.9} & \textbf{51.8} & \textbf{75.2} & \textbf{45.5} \\
    SBS(5,3)+\textsc{MASPRM}  & \textbf{77.5} & \textbf{51.3} & \textbf{74.8} & \textbf{45.0} \\

    \bottomrule
\end{tabular}
  }
\end{table}

\noindent
Guided MAS search with 1.5B agents and \textsc{MASPRM-7B} substantially improves over the direct \texttt{Qwen2.5-1.5B-Instruct} baseline on every dataset.
MCTS(10,3)+\textsc{MASPRM} gains +41.0, +34.3, +40.9, and +19.7 points on \textsc{GSM8K}, \textsc{MATH}, \textsc{MMLU}, and \textsc{LogiQA}, respectively; SBS(5,3)+\textsc{MASPRM} gains +35.6, +33.8, +40.5, and +19.2 points.
Relative to the larger direct single-agent baselines listed here, guided MAS is strongest on \textsc{GSM8K} and \textsc{MMLU} and roughly competitive on \textsc{MATH} (51.8 vs. 51.9 for \texttt{Mistral~24B-Instruct}).
On \textsc{LogiQA}, however, MCTS(10,3)+\textsc{MASPRM} reaches 45.5, below \texttt{GPT-4o-mini} at 50.1 and \texttt{Mistral~24B-Instruct} at 56.4.

\subsection{\texorpdfstring{Hit@$k$ sensitivity under limited candidates}{Hit@k sensitivity under limited candidates}}
\label{app:hitk-sensitivity}
We study robustness when only a small number of candidate solutions can be generated and returned. Table~\ref{tab:masprm-hitk} in App.~\ref{app:hitk-sensitivity} reports Hit@$k$ for $k \in \{1,3,5\}$ under MCTS(10,3), SC@5, and SBS(5,3) with \texttt{Qwen2.5-1.5B-Instruct} agents. For SBS(5,3), only three terminal candidates are returned, so Hit@5 is not applicable. Policy likelihood scoring shows Hit@1 to Hit@5 gaps averaging +16.1 points for MCTS and +19.0 points for SC@5. \textsc{MASPRM} scoring reduces these gaps to +5.8 points (MCTS) and +10.5 points (SC@5), improvements of 10.3 and 8.5 points respectively.

\begin{table*}[!htbp]
  \centering
  \scriptsize
  \setlength{\tabcolsep}{3pt}
  \caption{Hit@$k$ (\%) for $k\in\{1,3,5\}$ under MCTS(10,3), SC@5, and SBS(5,3).}
  \label{tab:masprm-hitk}
  \resizebox{\textwidth}{!}{
  \begin{tabular}{l
                  ccc    
                  ccc    
                  ccc    
                  ccc}   
    \toprule
    \multirow{2}{*}{Method} &
    \multicolumn{3}{c}{\textsc{GSM8K}} &
    \multicolumn{3}{c}{\textsc{MATH}} &
    \multicolumn{3}{c}{\textsc{MMLU}} &
    \multicolumn{3}{c}{\textsc{LogiQA}} \\
    \cmidrule(lr){2-4}
    \cmidrule(lr){5-7}
    \cmidrule(lr){8-10}
    \cmidrule(lr){11-13}
      & Hit@1 & Hit@3 & Hit@5 &
        Hit@1 & Hit@3 & Hit@5 &
        Hit@1 & Hit@3 & Hit@5 &
        Hit@1 & Hit@3 & Hit@5 \\
    \midrule
    MCTS(10,3) (policy likelihood) &
        62.5 & 76.4 & 82.1 &
        42.5 & 54.5 & 60.6 &
        60.5 & 68.2 & 71.3 &
        36.5 & 47.3 & 52.4 \\
    \rowcolor{MASPRMRowBlue}
    MCTS(10,3) (\textsc{MASPRM-7B}) &
        82.9 & 85.0 & 88.2 &
        51.8 & 57.5 & 60.5 &
        75.2 & 76.7 & 77.4 &
        45.5 & 50.5 & 52.5 \\
    SC@5 (policy likelihood) &
        61.2 & 73.3 & 81.4 &
        41.5 & 50.6 & 57.5 &
        54.5 & 69.2 & 74.9 &
        35.9 & 46.5 & 55.4 \\
    \rowcolor{MASPRMRowBlue}
    SC@5 (\textsc{MASPRM-7B}) &
        67.8 & 77.0 & 81.4 &
        49.2 & 55.3 & 57.5 &
        67.2 & 72.7 & 74.9 &
        43.0 & 50.2 & 55.2 \\
    SBS(5,3) (policy likelihood) &
        60.3 & 66.8 & -- &
        40.4 & 45.6 & -- &
        59.4 & 59.7 & -- &
        37.0 & 40.5 & -- \\
    \rowcolor{MASPRMRowBlue}
    SBS(5,3) (\textsc{MASPRM-7B}) &
        77.5 & 79.5 & -- &
        51.3 & 55.8 & -- &
        74.8 & 77.6 & -- &
        45.0 & 47.7 & -- \\
    \bottomrule
  \end{tabular}
  }
\end{table*}

\subsection{Cross-LLM transfer results}
\label{app:cross-llm}

We test whether reward models trained on \texttt{Qwen2.5-1.5B-Instruct} rollouts can guide a different agent LLM. We replace only the agent model with \texttt{Llama3.2~3B-Instruct}. The reward models remain fixed and are not retrained. We use the same inference procedures, metrics, and budget accounting as Sec.~\ref{sec:exp-setup}.

\begin{table*}[!htbp]
  \centering
  \scriptsize
  \setlength{\tabcolsep}{3pt}
  \caption{Cross-LLM results with \texttt{Llama3.2~3B-Instruct} agents and fixed \texttt{Qwen2.5}-trained reward models (no retraining): Hit@1 (\%), TOK ($\times 10^3$), and AC.}
  \label{tab:main-results-masprm-cross-llm}
  \resizebox{\textwidth}{!}{
  \begin{tabular}{ll
                  c >{\tiny}c >{\tiny}c    
                  c >{\tiny}c >{\tiny}c    
                  c >{\tiny}c >{\tiny}c    
                  c >{\tiny}c >{\tiny}c}   
    \toprule
    \multirow{2}{*}{Search} &
    \multirow{2}{*}{Scorer} &
    \multicolumn{3}{c}{\textsc{GSM8K}} &
    \multicolumn{3}{c}{\textsc{MATH}} &
    \multicolumn{3}{c}{\textsc{MMLU}} &
    \multicolumn{3}{c}{\textsc{LogiQA}} \\
    \cmidrule(lr){3-5}
    \cmidrule(lr){6-8}
    \cmidrule(lr){9-11}
    \cmidrule(lr){12-14}
      & &
      Hit@1 & TOK & AC &
      Hit@1 & TOK & AC &
      Hit@1 & TOK & AC &
      Hit@1 & TOK & AC \\
    \midrule

    Single-Pass & -- &
      52.6 & 0.8 & 4 &
      35.2 & 1.2 & 4 &
      56.0 & 0.1 & 3 &
      36.3 & 0.5 & 3 \\

    \midrule

    SC@5 & Policy Likelihood &
      64.2 & 3.7 & 20 &
      44.1 & 5.6 & 20 &
      56.4 & 0.3 & 15 &
      37.2 & 2.6 & 15 \\

    SC@5 & ORM-7B &
      68.8 & 3.7 & 20 &
      47.8 & 5.6 & 20 &
      63.4 & 0.3 & 15 &
      42.1 & 2.6 & 15 \\

    \rowcolor{MASPRMRowBlue}
    SC@5 & \textsc{MASPRM-7B} &
      \textbf{69.8} & 3.7 & 20 &
      \textbf{48.6} & 5.6 & 20 &
      \textbf{66.7} & 0.3 & 15 &
      \textbf{43.3} & 2.6 & 15 \\

    \midrule
    SBS(5,1) & Policy Likelihood &
      58.4 & 3.8 & 20 &
      42.1 & 5.5 & 20 &
      55.0 & 0.3 & 15 &
      36.1 & 2.5 & 15 \\

    \rowcolor{MASPRMRowBlue}
    SBS(5,1) & \textsc{MASPRM-7B} &
      \textbf{74.6} & 3.7 & 20 &
      \textbf{49.1} & 5.5 & 20 &
      \textbf{71.6} & 0.3 & 15 &
      \textbf{44.1} & 2.5 & 15 \\

    \midrule
    SBS(5,3) & Policy Likelihood &
      63.1 & 8.3 & 50 &
      41.3 & 12.3 & 50 &
      58.5 & 0.8 & 35 &
      37.6 & 6.5 & 35 \\

    \rowcolor{MASPRMRowBlue}
    SBS(5,3) & \textsc{MASPRM-7B} &
      \textbf{78.2} & 8.5 & 50 &
      \textbf{52.2} & 12.5 & 50 &
      \textbf{74.3} & 0.8 & 35 &
      \textbf{45.4} & 6.5 & 35 \\

    \midrule
    MCTS(10,3) & Policy Likelihood &
      64.1 & 5.9 & 38.7 &
      43.4 & 9.8 & 34.5 &
      61.9 & 0.2 & 12.8 &
      37.4 & 2.5 & 13.9 \\


    MCTS(10,3) & ORM-7B &
      67.3 & 5.7 & 35.8 &
      46.9 & 9.7 & 34.2 &
      67.7 & 0.2 & 12.8 &
      42.1 & 2.5 & 13.8 \\

    \rowcolor{MASPRMRowBlue}
    MCTS(10,3) & \textsc{MASPRM-7B} &
      \textbf{75.3} & 5.4 & 35.4 &
      \textbf{52.1} & 9.7 & 37.8 &
      \textbf{74.9} & 0.3 & 13.1 &
      \textbf{46.1} & 2.6 & 14.0 \\

    \bottomrule
  \end{tabular}
  }
\end{table*}

\noindent
Across SC@5, SBS, and MCTS, \textsc{MASPRM-7B} gives the highest Hit@1 in every dataset. Under MCTS(10,3), gains over policy likelihood are +11.2 points on \textsc{GSM8K}, +8.7 on \textsc{MATH}, +13.0 on \textsc{MMLU}, and +8.7 on \textsc{LogiQA}. Gains over ORM-7B are +8.0, +5.2, +7.2, and +4.0 points.

\subsection{Head-type ablation for \textsc{MASPRM} and ORM}
\label{app:head-ablation}

We ablate reward-model head choices under a fixed search budget. We focus on MCTS with $N_{\mathrm{sim}}{=}10$ simulations (MCTS(10,3)) using \texttt{Qwen2.5-1.5B-Instruct} as the agent LLM and compare:

ORM variants trained with BCE vs.\ MSE heads.

\textsc{MASPRM-1.5B} and \textsc{MASPRM-7B} with MSE vs.\ BT heads.

We report Hit@1 on all four datasets in Table~\ref{tab:masprm-head-ablation} in App.~\ref{app:head-ablation}.

For \textsc{MASPRM}, we use the same MAS-MCTS rollouts and preprocessing as App.~\ref{app:masprm-data}. For the BT head, we train on extracted preference pairs with the Bradley-Terry ranking loss (Sec.~\ref{sec:masprm-training}). For the MSE head, we train pointwise by minimizing $\ell_{\mathrm{MSE}}=\big(V_\phi(s')-y_{\mathrm{proc}}(s')\big)^2$ to the clipped process targets.

For ORM training, we discard intermediate states and use only terminal outcomes from the same rollouts:
$\mathcal{D}_{\mathrm{out}}=\{(s_T, r)\}$ with $r=R(s_T)$.
The BCE ORM treats correctness as a binary label, while the MSE ORM regresses to the scalar reward $r\in\{-1,+1\}$.

\begin{table}[!htbp]
  \centering
  \scriptsize
  \setlength{\tabcolsep}{6pt}
  \caption{Head-type ablation (Hit@1, \%) under MCTS(10,3) with \texttt{Qwen2.5-1.5B-Instruct} agents.}
  \label{tab:masprm-head-ablation}
  \resizebox{\linewidth}{!}{
  \begin{tabular}{lcccc}
    \toprule
    Head &
    \textsc{GSM8K} &
    \textsc{MATH}  &
    \textsc{MMLU} &
    \textsc{LogiQA} \\
    \midrule
    \multicolumn{5}{l}{\textbf{ORM-1.5B}} \\
    \rowcolor{MASPRMRowBlue}
    BCE                 & 66.8 & 45.3 & 66.2 & 40.2 \\
    MSE                 & 66.1 & 44.5 & 65.4 & 39.6 \\
    \midrule
    \multicolumn{5}{l}{\textbf{ORM-7B}} \\
    \rowcolor{MASPRMRowBlue}
    BCE                 & 68.4 & 46.2 & 67.1 & 41.4 \\
    MSE                 & 68.2 & 46.2 & 66.3 & 40.8 \\
    \midrule
    \multicolumn{5}{l}{\textbf{\textsc{MASPRM-1.5B}}} \\
    \rowcolor{MASPRMRowBlue}
    BT         & 69.8 & 47.3 & 68.8 & 42.2 \\
    MSE        & 65.5 & 44.1 & 67.6 & 41.9 \\
    \midrule
    \multicolumn{5}{l}{\textbf{\textsc{MASPRM-7B}}} \\
    \rowcolor{MASPRMRowBlue}
    BT           & 82.9 & 51.8 & 75.2 & 45.5 \\
    MSE          & 73.3 & 50.6 & 74.9 & 44.4 \\
    \bottomrule
  \end{tabular}
  }
\end{table}

\noindent
BT is the best \textsc{MASPRM} head on all four datasets for both model sizes. Gains over MSE are +0.3 to +4.3 points for the 1.5B model and +0.3 to +9.6 points for the 7B model. For ORM, BCE is best or tied: gains over MSE are +0.6 to +0.8 points for the 1.5B model and 0.0 to +0.8 points for the 7B model. These results support using BT for \textsc{MASPRM} and BCE for ORM in the main comparison.

\subsection{Local observability ablation for \textsc{MASPRM}}
\label{app:local-observability}

We compare the default \textsc{MASPRM} scorer, which receives the full routed transcript, to a local-observability variant. At each scoring step, the local variant receives only the task input and the acting agent's local routed view, including that agent's candidate output, rather than the full routed transcript.

The setup is controlled: both variants use the same rollouts, the same \texttt{Qwen2.5-1.5B-Instruct} agent model, the same search budget, the same MCTS(10,3) setting, and the same \textsc{MASPRM-7B} backbone as the main MCTS results. Table~\ref{tab:masprm-local-observability} in App.~\ref{app:local-observability} reports Hit@1 on all four datasets.

\begin{table}[!htbp]
  \centering
  \scriptsize
  \setlength{\tabcolsep}{6pt}
  \caption{Hit@1 under MCTS(10,3) with \texttt{Qwen2.5-1.5B-Instruct} agents, comparing full vs.\ local scorer observability.}
  \label{tab:masprm-local-observability}
  \resizebox{\linewidth}{!}{
  \begin{tabular}{lcccc}
    \toprule
    Scorer view &
    \textsc{GSM8K} &
    \textsc{MATH}  &
    \textsc{MMLU} &
    \textsc{LogiQA} \\
    \midrule
    \rowcolor{MASPRMRowBlue}
    Full routed transcript & \textbf{82.9} & \textbf{51.8} & \textbf{75.2} & \textbf{45.5} \\
    Local agent view       & 77.4 & 49.4 & 71.7 & 43.3 \\
    \bottomrule
  \end{tabular}
  }
\end{table}

\noindent
The full routed transcript gives the highest Hit@1 on all four datasets. The local view lowers Hit@1 by 5.5 points on \textsc{GSM8K}, 2.4 on \textsc{MATH}, 3.5 on \textsc{MMLU}, and 2.2 on \textsc{LogiQA}, while remaining above the policy likelihood MCTS baselines in Table~\ref{tab:main-results-masprm-qwen15b}.

\subsection{\texorpdfstring{\textsc{DyLAN}-style dynamic scheduler on \textsc{GSM8K}}{DyLAN-style dynamic scheduler on GSM8K}}
\label{app:dylan-dynamic}

We include a separate GSM8K dynamic-scheduler experiment inspired by \textsc{DyLAN}~\cite{liu2024dylan}. This experiment is not part of the graph-sensitivity ablation in App.~\ref{app:graph-sensitivity}. The dynamic setup uses three solver agents, a judge that checks agreement, and a conditional refinement layer. If the judge emits \texttt{VERDICT: AGREE}, the judge output is used as the terminal answer. If it emits \texttt{VERDICT: DISAGREE}, three refiners run and a final judge returns the terminal answer. We express this as an acyclic graph with conditional edges so the same MAS-MCTS data generation and inference code can be used.

Table~\ref{tab:gsm8k-dylan-dynamic} reports Single-Pass and MCTS(10,3) for both setups. All rows use \textsc{GSM8K} and \texttt{Qwen2.5-1.5B-Instruct} generator agents. The Single-Pass rows use no scorer. The MCTS rows use \textsc{MASPRM-7B}. For each setup, training data generation and testing are run from scratch: we generate MAS-MCTS(40,3) rollouts on the GSM8K training split for that setup, train a new \textsc{MASPRM-7B} on those rollouts, and evaluate MCTS(10,3) on the GSM8K test split. Other settings follow App.~\ref{app:exp-details}. The exact prompts and conditional edges are in App.~\ref{app:graph}.

\begin{table*}[!htbp]
  \centering
  \scriptsize
  \caption{\textsc{GSM8K} dynamic-scheduler feasibility check with \texttt{Qwen2.5-1.5B-Instruct} generator agents. We report this as preliminary evidence that \textsc{MASPRM} extends to conditional schedules, not as a head-to-head comparison with MAS$_1$. The MCTS(10,3) rows use \textsc{MASPRM-7B}; each setup uses fresh MAS-MCTS training data and a newly trained scorer.}
  \label{tab:gsm8k-dylan-dynamic}
  \resizebox{\linewidth}{!}{
  \begin{tabular}{lllccc}
    \toprule
    Setup & Search & Scorer & Hit@1 (\%) & TOK ($\times 10^3$) & AC \\
    \midrule
    MAS$_1$ & Single-Pass & -- & \meanstd{47.2}{1.9} & \meanstd{0.7}{0.0} & 4 \\
    \textsc{DyLAN}-style dynamic MAS & Single-Pass & -- &\meanstd{51.2}{1.9} & \meanstd{1.1}{0.5} & \meanstd{5.8}{0.9} \\
    MAS$_1$ & MCTS(10,3) & \textsc{MASPRM-7B} & \meanstd{82.9}{1.2} & \meanstd{5.5}{0.1} & \meanstd{34.4}{1.7} \\
    \textsc{DyLAN}-style dynamic MAS & MCTS(10,3) & \textsc{MASPRM-7B} & \meanstd{84.6}{2.1} & \meanstd{11.9}{4.4} & \meanstd{93.8}{1.2} \\
    \bottomrule
  \end{tabular}
  }
\end{table*}

\subsection{Graph sensitivity ablation on \textsc{GSM8K}}
\label{app:graph-sensitivity}

We evaluate three GSM8K MAS instantiations, denoted MAS$_1$-MAS$_3$.
All three share the same generator model family and reward backbone: the agents are \texttt{Qwen2.5-1.5B-Instruct}, and the scorer is \textsc{MASPRM-7B} built on \texttt{Qwen2.5-7B}.
The scorer is trained on MAS$_1$ rollouts and then transferred unchanged to MAS$_2$ and MAS$_3$.
We keep the search/decoding hyperparameters fixed (App.~\ref{app:exp-details}).
The only differences across MAS$_1$-MAS$_3$ are the agent role prompts and the directed communication graph (which determines what each agent can condition on at its turn).

MAS definitions are as follows. MAS$_1$ corresponds to a plan-then-solve pipeline with verification (Reader $\rightarrow$ Planner $\rightarrow$ Solver $\rightarrow$ Verifier) \cite{wang2023plan}.
MAS$_2$ corresponds to the critique and revise pipeline (Solver $\rightarrow$ Evaluator $\rightarrow$ Reflector $\rightarrow$ Reviser) \cite{madaan2023self}.
MAS$_3$ corresponds to a consensus style pipeline with two independent solutions and a final judge (Reader $\rightarrow$ Solver$_A$ $\rightarrow$ Solver$_B$ $\rightarrow$ Judge), inspired by self-consistency \cite{wang2023selfconsistency}.
Exact prompts and graphs are in App.~\ref{app:graph}.

Under the \textsc{MASPRM} scoring protocol, at a fixed prefix state $s$, candidate continuations share identical previous steps and differ only in the next newly generated step.
Concretely, for candidate next step messages $a$, we form successor states $s'=\mathrm{next}(s,a)$ and score each $s'$ with \textsc{MASPRM}. Search selects and expands based on these relative scores.

\begin{table}[!htbp]
  \centering
  \scriptsize
  \setlength{\tabcolsep}{6pt}
  \caption{\textsc{GSM8K} graph sensitivity (Hit@1, \%) across MAS$_1$-MAS$_3$.}
  \label{tab:gsm8k-graph-sensitivity}
  \resizebox{\linewidth}{!}{
  \begin{tabular}{lccc}
    \toprule
    Method & MAS$_1$ & MAS$_2$ & MAS$_3$ \\
    \midrule
    Single-Pass & 47.2 & 54.4 & 56.2 \\
    SBS(5,1)+\textsc{MASPRM} & 73.4 & 73.7 & 75.2 \\
    SBS(5,3)+\textsc{MASPRM} & 77.5 & 78.5 & 79.7 \\
    MCTS(10,3)+\textsc{MASPRM} & 82.9 & 81.4 & 80.9 \\
    \bottomrule
\end{tabular}
  }
\end{table}

\noindent Single-Pass performance varies by 9.0 points across MAS$_1$ to MAS$_3$ (47.2 to 56.2). With \textsc{MASPRM} guidance, all three configurations are within 1.8 to 2.2 points (SBS/MCTS ranges: 73.4 to 75.2, 77.5 to 79.7, 80.9 to 82.9), reducing the ranges by 6.8 to 7.2 points.

\subsection{OOD transfer protocol}
\label{app:ood-transfer-protocol}
For Table~\ref{tab:masprm-ood-with-logprob}, we train \textsc{MASPRM-7B} on one source dataset and evaluate it on a held out target dataset without further tuning. We use MCTS(10,3), \texttt{Qwen2.5-1.5B-Instruct} agents, and the same answer extraction as the main experiments. We evaluate within-domain transfers: \textsc{GSM8K} to \textsc{MATH}, \textsc{MATH} to \textsc{GSM8K}, \textsc{MMLU} to \textsc{LogiQA}, and \textsc{LogiQA} to \textsc{MMLU}. The policy likelihood row is the no-reward-model baseline under the same search setting.

Cross-dataset transfer stays below the in-domain diagonal entries, but remains above policy likelihood on each target by +5.3 to +11.1 points.

\subsection{Wall-clock Comparison}
\label{app:wall-clock}
We measure wall-clock time on \textsc{GSM8K} using the hardware in App.~\ref{app:hardware}. An AC is one agent generation, and an RC is one \textsc{MASPRM} scorer forward pass. The mean AC time is 18.94\,s, while the mean RC time is 0.16\,s.

Estimated per-instance wall-clock on \textsc{GSM8K}: MCTS(10,3) $\approx$ 11\,min; SBS(5,3) $\approx$ 16\,min. In this setup, one RC is about 118 times faster than one AC.

\section{MAS Graph, Prompts, and Configuration}
\label{app:graph}

This work uses multiple MAS instantiations (varying in depth $D$, agent roles, and directed communication graphs) rather than a single fixed MAS for all settings. The overall design choices are inspired by: decomposition/plan-then-solve prompting \cite{wang2023plan}, self-refinement/self-critique loops \cite{madaan2023self}, and multi-agent debate \cite{du2024improving,liang2024encouraging}. Below we list the exact YAML-style configurations used, plus the GSM8K MAS$_3$ consensus configuration introduced for the graph sensitivity ablation and the GSM8K \textsc{DyLAN}-style dynamic scheduler used in App.~\ref{app:dylan-dynamic}.
In these implementation-style listings, agent IDs are 0-indexed, and the integer $-1$ is an alias for the question node $\text{q}$ from Sec.~\ref{sec:preliminaries}.

\subsection{Agent roles, system prompts, and decoding}
Each MAS is specified by a list of agents (name, system prompt, and a per-agent decoding cap). For readability, we present each MAS's \texttt{agents} block.

\paragraph{MAS$_{\text{Single}}$: single-agent baseline.}
\begin{tcolorbox}[breakable, width=\linewidth, colback=gray!5, colframe=gray!40!black, left=2mm, right=2mm, boxsep=1mm]
\begin{lstlisting}
- name: Assistant
system_prompt: |
  You are an intelligent assistant that helps users by providing accurate and concise answers.
  Solve the problem carefully.
  Output exactly:
  Final Answer: <your answer>
max_new_tokens: 1024
\end{lstlisting}
\end{tcolorbox}
This is a standard single-pass baseline aligned with chain-of-thought style prompting \cite{wei2022chain}.

\paragraph{MAS$_{\text{RPSV}}$ (GSM8K MAS$_1$/MATH): Reader $\rightarrow$ Planner $\rightarrow$ Solver $\rightarrow$ Verifier.}
\begin{tcolorbox}[breakable, width=\linewidth, colback=gray!5, colframe=gray!40!black, left=2mm, right=2mm, boxsep=1mm]
\begin{lstlisting}
- name: Reader
system_prompt: "You are the Reader. Extract key facts and restate the question."
max_new_tokens: 1024
- name: Planner
system_prompt: "You are the Planner. Propose a concise plan to solve the task."
max_new_tokens: 1024
- name: Solver
system_prompt: "You are the Solver. Carry out the plan and compute results."
max_new_tokens: 1024
- name: Verifier
system_prompt: "You are the Verifier. Double-check the result and produce the final answer only."
max_new_tokens: 1024
\end{lstlisting}
\end{tcolorbox}
This is a plan-then-solve decomposition with an explicit verification stage \cite{wang2023plan}.

\paragraph{MAS$_{\text{Refine}}$ (GSM8K MAS$_2$): Solver $\rightarrow$ Evaluator $\rightarrow$ Reflector $\rightarrow$ Reviser.}
\begin{tcolorbox}[breakable, width=\linewidth, colback=gray!5, colframe=gray!40!black, left=2mm, right=2mm, boxsep=1mm]
\begin{lstlisting}
- name: Agent_Solver
system_prompt: |
  You are a math solver. Solve the problem step by step.

  First show your work, then give the final answer.
  End with: FINAL ANSWER: [your answer]
max_new_tokens: 512

- name: Agent_Evaluator
system_prompt: |
  You are a checker. Look at the solution and check if it is correct.

  Check:
  - Are the steps logical?
  - Is the math correct?

  Reply with only one of these:
  CORRECT
  or
  INCORRECT: [explain the error]
max_new_tokens: 256

- name: Agent_Reflector
system_prompt: |
  You are an advisor. A solution was wrong.

  Give one short tip to fix the mistake.
  Do not solve the problem yourself.

  Reply with: TIP: [your advice]
max_new_tokens: 256

- name: Agent_Reviser
system_prompt: |
  You are a math solver. A previous answer was wrong.

  Use the tip given to solve the problem correctly.
  Show your work step by step.
  End with: FINAL ANSWER: [your answer]
max_new_tokens: 512
\end{lstlisting}
\end{tcolorbox}
This configuration implements an explicit feedback$\rightarrow$refine loop \cite{madaan2023self}.

\paragraph{MAS$_{\text{Consensus}}$ (GSM8K MAS$_3$): Reader $\rightarrow$ Solver$_A$ $\rightarrow$ Solver$_B$ $\rightarrow$ Judge.}
\begin{tcolorbox}[breakable, width=\linewidth, colback=gray!5, colframe=gray!40!black, left=2mm, right=2mm, boxsep=1mm]
\begin{lstlisting}
- name: Reader
system_prompt: "You are the Reader. Extract key facts and restate the question."
max_new_tokens: 512

- name: Solver_A
system_prompt: |
  You are Solver A. Solve the problem step by step.
  Be careful with arithmetic.
  End with: FINAL ANSWER: <number>
max_new_tokens: 1024

- name: Solver_B
system_prompt: |
  You are Solver B. Independently solve the same problem from scratch.
  Try a different approach if possible and be careful with arithmetic.
  End with: FINAL ANSWER: <number>
max_new_tokens: 1024

- name: Judge
system_prompt: |
  You are the Judge/Verifier. You receive:
  (i) the original question,
  (ii) the Reader summary,
  (iii) Solver_A's solution,
  (iv) Solver_B's solution.

  If both solvers agree and the reasoning is sound, output exactly:
  Final Answer: <value>

  If they disagree or either contains an error, identify the mistake and recompute cleanly.
  Then output exactly:
  Final Answer: <value>
max_new_tokens: 512
\end{lstlisting}
\end{tcolorbox}
This configuration is motivated by self-consistency style aggregation and final answer selection \cite{wang2023selfconsistency}.

\paragraph{MAS$_{\text{DyLAN}}$ (GSM8K dynamic scheduler): solver pool $\rightarrow$ Judge $\rightarrow$ optional refiners $\rightarrow$ FinalJudge.}
\begin{tcolorbox}[breakable, width=\linewidth, colback=gray!5, colframe=gray!40!black, left=2mm, right=2mm, boxsep=1mm]
\begin{lstlisting}
- name: SolverA_MathAnalyst
system_prompt: |
  You are a careful Math Analyst. Read the problem, list the relevant
  quantities and relationships, then solve step by step, showing
  arithmetic. End with a single line in the exact format:
  Final Answer: <number>
max_new_tokens: 512

- name: SolverB_AlgebraSolver
system_prompt: |
  You are an Algebra Solver. Set up explicit variables and equations,
  then solve them step by step, showing arithmetic. End with a single
  line in the exact format:
  Final Answer: <number>
max_new_tokens: 512

- name: SolverC_ArithmeticReasoner
system_prompt: |
  You are an Arithmetic Reasoner. Translate the problem into a
  sequence of concrete arithmetic operations and execute each one,
  showing your work. End with a single line in the exact format:
  Final Answer: <number>
max_new_tokens: 512

- name: Judge
system_prompt: |
  You are the Judge. You receive the original question and three
  candidate solutions from peer solvers. Independently re-derive the
  answer and compare it against the candidates.

  If at least two of the three candidates agree with your re-derived
  answer AND the reasoning is sound, output EXACTLY:
  VERDICT: AGREE
  Final Answer: <number>

  Otherwise, output EXACTLY:
  VERDICT: DISAGREE
  <one short paragraph identifying where the candidates went wrong;
  do NOT give a final number on the second line>
max_new_tokens: 384

- name: RefinerA
system_prompt: |
  You are Refiner A. You receive the original question, your own
  previous draft, and the Judge's critique. Produce a corrected
  solution that addresses the critique, showing arithmetic. End with
  a single line in the exact format:
  Final Answer: <number>
max_new_tokens: 512

- name: RefinerB
system_prompt: |
  You are Refiner B. You receive the original question, your own
  previous draft, and the Judge's critique. Produce a corrected
  solution that addresses the critique, showing arithmetic. End with
  a single line in the exact format:
  Final Answer: <number>
max_new_tokens: 512

- name: RefinerC
system_prompt: |
  You are Refiner C. You receive the original question, your own
  previous draft, and the Judge's critique. Produce a corrected
  solution that addresses the critique, showing arithmetic. End with
  a single line in the exact format:
  Final Answer: <number>
max_new_tokens: 512

- name: FinalJudge
system_prompt: |
  You are the Final Judge. You receive the original question and
  three refined solutions. Decide on the correct numeric answer
  (re-derive if needed) and output EXACTLY one line:
  Final Answer: <number>
max_new_tokens: 256
\end{lstlisting}
\end{tcolorbox}
This configuration follows the \textsc{DyLAN} idea of diverse solvers, agent communication, and a dynamic scheduler, implemented here as a judge-gated refinement layer \cite{liu2024dylan}.

\paragraph{MAS$_{\text{Debate}}$: Angel/Devil debate with Judge aggregation (MMLU).}
\begin{tcolorbox}[breakable, width=\linewidth, colback=gray!5, colframe=gray!40!black, left=2mm, right=2mm, boxsep=1mm]
\begin{lstlisting}
- name: Angel
system_prompt: |
  You are a helpful debater. Read the multiple-choice question with options A-D.
  Reason briefly and choose the best option.
max_new_tokens: 1024

- name: Devil
system_prompt: |
  You are a skeptical debater. Read the same question.
  Challenge assumptions and choose the best option.
max_new_tokens: 1024

- name: Judge
system_prompt: |
  You are the Judge. Read the question and both debaters' letters.
  Decide the final answer.
  Respond EXACTLY with this format:
  Final Answer: X
  where X is one of A, B, C, or D.
max_new_tokens: 1024
\end{lstlisting}
\end{tcolorbox}
This configuration follows debate and judge style aggregation \cite{liang2024encouraging}.

\paragraph{MAS$_{\text{Logic}}$: constraint extraction $\rightarrow$ world building $\rightarrow$ option checking (LogiQA).}
\begin{tcolorbox}[breakable, width=\linewidth, colback=gray!5, colframe=gray!40!black, left=2mm, right=2mm, boxsep=1mm]
\begin{lstlisting}
- name: Constraint_Extractor
system_prompt: |
  You are the Constraint Extractor.
  Ignore the answer options for now.
  Extract all explicit constraints and logical rules stated in the passage.
  Format them as a numbered list (e.g., "1) A != B", "2) If C then D").
max_new_tokens: 1024

- name: World_Builder
system_prompt: |
  You are the World Builder.
  You receive a list of constraints.
  Construct one or more concrete "valid states" that satisfy all constraints.
  If multiple states are possible, describe what must be true in all valid states.
max_new_tokens: 1024

- name: Query_Checker
system_prompt: |
  You are the Query Checker.
  You receive the question (with options) and the valid state summary.
  Determine which option is entailed by (or most consistent with) the valid state(s).
  Output exactly:
  Final Answer: X
  where X is one of A, B, C, or D.
max_new_tokens: 1024
\end{lstlisting}
\end{tcolorbox}
This configuration uses decomposition into structured intermediate artifacts \cite{zhou2023leasttomost}.

\subsection{Communication edges and schedule}
Edges are directed. The integer \(-1\) follows the implementation-style alias above. For fixed MAS, the schedule order is one turn per agent in the order listed in the \texttt{agents} block. The \textsc{DyLAN}-style dynamic MAS follows the listed topological order and conditionally enables the refinement layer based on the Judge output.
Edges to the terminal node \(\mathrm{sink}\) from the active terminal agent are implicit and omitted from the YAML-style edge lists.

\paragraph{MAS$_{\text{Single}}$ schedule.}
Schedule order: \texttt{[Assistant]}. \\
Edges:
\begin{tcolorbox}[breakable, title=Edges, colback=gray!5, colframe=gray!40!black]
\begin{lstlisting}
edges:
  - [-1, 0]
\end{lstlisting}
\end{tcolorbox}

\paragraph{MAS$_{\text{RPSV}}$ schedule.}
Schedule order: \texttt{[Reader, Planner, Solver, Verifier]}. \\
Edges:
\begin{tcolorbox}[breakable, title=Edges, colback=gray!5, colframe=gray!40!black]
\begin{lstlisting}
edges:
  - [-1, 0]
  - [-1, 1]
  - [0, 1]
  - [1, 2]
  - [-1, 3]
  - [2, 3]
\end{lstlisting}
\end{tcolorbox}

\paragraph{MAS$_{\text{Refine}}$ schedule.}
Schedule order: \texttt{[Agent\_Solver, Agent\_Evaluator, Agent\_Reflector, Agent\_Reviser]}. \\
Edges:
\begin{tcolorbox}[breakable, title=Edges, colback=gray!5, colframe=gray!40!black]
\begin{lstlisting}
edges:
  - [-1, 0] 
  - [0, 1] 
  - [0, 2]  
  - [1, 2]  
  - [-1, 3] 
  - [2, 3]  
\end{lstlisting}
\end{tcolorbox}

\paragraph{MAS$_{\text{Consensus}}$ schedule.}
Schedule order: \texttt{[Reader, Solver\_A, Solver\_B, Judge]}. \\
Edges:
\begin{tcolorbox}[breakable, title=Edges, colback=gray!5, colframe=gray!40!black]
\begin{lstlisting}
edges:
  - [-1, 0]  
  - [-1, 1] 
  - [-1, 2] 
  - [ 0, 1]  
  - [ 0, 2]  
  - [-1, 3] 
  - [ 0, 3]  
  - [ 1, 3]  
  - [ 2, 3] 
\end{lstlisting}
\end{tcolorbox}

\paragraph{MAS$_{\text{DyLAN}}$ dynamic schedule.}
Topological order: \texttt{[SolverA\_MathAnalyst, SolverB\_AlgebraSolver, SolverC\_ArithmeticReasoner, Judge, RefinerA, RefinerB, RefinerC, FinalJudge]}. The conditional edges into the refiners are enabled only when the Judge emits \texttt{VERDICT: DISAGREE}. If the Judge emits \texttt{VERDICT: AGREE}, the refinement layer is skipped and the Judge output is treated as the terminal answer. \\
Edges:
\begin{tcolorbox}[breakable, title=Edges, colback=gray!5, colframe=gray!40!black]
\begin{lstlisting}
edges:
  - [-1, 0]
  - [-1, 1]
  - [-1, 2]
  - [-1, 3]
  - [ 0, 3]
  - [ 1, 3]
  - [ 2, 3]
  - [-1, 4]
  - [ 0, 4]
  - [ 3, 4, "regex:(?i)verdict:\\s*disagree"]
  - [-1, 5]
  - [ 1, 5]
  - [ 3, 5, "regex:(?i)verdict:\\s*disagree"]
  - [-1, 6]
  - [ 2, 6]
  - [ 3, 6, "regex:(?i)verdict:\\s*disagree"]
  - [-1, 7]
  - [ 4, 7]
  - [ 5, 7]
  - [ 6, 7]
\end{lstlisting}
\end{tcolorbox}

\paragraph{MAS$_{\text{Debate}}$ schedule.}
Schedule order: \texttt{[Angel, Devil, Judge]}. \\
Edges:
\begin{tcolorbox}[breakable, title=Edges, colback=gray!5, colframe=gray!40!black]
\begin{lstlisting}
edges:
  - [-1, 0]
  - [-1, 1]
  - [-1, 2]
  - [0, 2]
  - [1, 2]
\end{lstlisting}
\end{tcolorbox}

\paragraph{MAS$_{\text{Logic}}$ schedule.}
Schedule order: \texttt{[Constraint\_Extractor, World\_Builder, Query\_Checker]}. \\
Edges:
\begin{tcolorbox}[breakable, title=Edges, colback=gray!5, colframe=gray!40!black]
\begin{lstlisting}
edges:
  - [-1, 0]
  - [0, 1]
  - [-1, 2]
  - [1, 2]
\end{lstlisting}
\end{tcolorbox}

\section{Algorithms (Brief Overview)}
\label{app:algorithms}

This appendix summarizes the procedures used for MAS-MCTS data generation during training, MCTS$(N_{\mathrm{sim}},C_{\max})$ with \textsc{MASPRM} during inference, and SBS$(B_2,B_1)$. The key difference between training and inference MCTS is the leaf evaluation: training backs up the ground truth terminal reward $R(s_T)$ (no virtual visits), whereas inference bootstraps with $V_\phi$ and initializes new children with virtual visits.
For the pseudocode, when in state $s=(x,H_{t-1},i_t)$, the acting agent is $i_t$ and $\pi^{(i_t)}(\cdot\mid x,H_{t-1}^{(i_t)})$ denotes its policy conditioned on its local view as defined in Sec.~\ref{sec:preliminaries}.

\begin{algorithm*}[t]
\caption{MAS-MCTS$(N_{\mathrm{sim}},C_{\max})$ Data Generation During Training}
\label{alg:train-mcts}
\begin{algorithmic}[1]
\Require Instance $(x, y^\star)$, graph $G$, schedule $\sigma$, simulations $N_{\mathrm{sim}}$, candidate cap $C_{\max}$, exploration $c_{\textsc{uct}}$
\State Initialize empty search tree with statistics $N(s,a)\!=\!0$, $W(s,a)\!=\!0$
\For{$k=1$ to $N_{\mathrm{sim}}$}
  \State $s \gets s_1(x)$
  \State $\text{path} \gets [\,]$
  \Comment{Selection}
  \While{$s$ is expanded and non-terminal}
    \State $a \gets \arg\max_{b\in \mathcal{A}^{\mathrm{exp}}_{i_t}(s)} \Big\{\widehat Q(s,b) + c_{\textsc{uct}}\sqrt{\frac{\ln(1+\sum_{u\in\mathcal{A}^{\mathrm{exp}}_{i_t}(s)}N(s,u))}{1+N(s,b)}}\Big\}$
    \State Append $(s,a)$ to $\text{path}$
    \State $s \gets \mathrm{next}(s,a)$
  \EndWhile
  \Comment{Expansion (no virtual visits in training)}
  \If{$s$ non-terminal}
    \State Sample up to $C_{\max}$ candidates $C_t$ from the acting agent's policy $\pi^{(i_t)}(\cdot\mid x,H_{t-1}^{(i_t)})$ where $s=(x,H_{t-1},i_t)$
    \State For each $a \in C_t$, create child $s'=\mathrm{next}(s,a)$
  \EndIf
  \Comment{Evaluation with ground truth (terminal reward)}
  \State Continue selection/expansion, appending each traversed edge to $\text{path}$, until reaching a terminal leaf $s_T$
  \State $v \gets R(s_T) \in \{-1, +1\}$
  \Comment{Backpropagation}
  \For{each $(u,b)$ in $\text{path}$}
    \State $N(u,b)\!\gets\!N(u,b)+1$
    \State $W(u,b)\!\gets\!W(u,b)+v$
    \State $\widehat Q(u,b)\!\gets\!\frac{W(u,b)}{N(u,b)}$
  \EndFor
\EndFor
\State Process targets: For every expanded edge $(s,a)$ with $N(s,a)>0$ and child $s'=\mathrm{next}(s,a)$, set $y_{\mathrm{proc}}(s') \gets \widehat Q(s,a)$ and add $(s', y_{\mathrm{proc}}(s'))$ to $\mathcal{D}_{\mathrm{proc}}$
\State Fit the process head: Within each parent state, form sibling preference pairs $(s^+,s^-)$ with $y_{\mathrm{proc}}(s^+)>y_{\mathrm{proc}}(s^-)$ and train $V_\phi$ with a Bradley-Terry ranking loss
\end{algorithmic}
\end{algorithm*}

\begin{algorithm*}[t]
\caption{MCTS$(N_{\mathrm{sim}},C_{\max})$ with \textsc{MASPRM} During Inference}
\label{alg:infer-mcts}
\begin{algorithmic}[1]
\Require Instance $x$, graph $G$, schedule $\sigma$, simulations $N_{\mathrm{sim}}$, candidate cap $C_{\max}$, exploration $c_{\textsc{uct}}$, value model $V_\phi$
\State Initialize empty search tree with statistics $N(s,a)\!=\!0$, $W(s,a)\!=\!0$
\For{$k=1$ to $N_{\mathrm{sim}}$}
  \State $s \gets s_1(x)$
  \State $\text{path} \gets [\,]$
  \Comment{Selection}
  \While{$s$ is expanded and non-terminal}
    \State $a \gets \arg\max_{b\in \mathcal{A}^{\mathrm{exp}}_{i_t}(s)} \Big\{\widehat Q(s,b) + c_{\textsc{uct}}\sqrt{\frac{\ln(1+\sum_{u\in\mathcal{A}^{\mathrm{exp}}_{i_t}(s)}N(s,u))}{1+N(s,b)}}\Big\}$
    \State Append $(s,a)$ to $\text{path}$
    \State $s \gets \mathrm{next}(s,a)$
  \EndWhile
  \Comment{Expansion with virtual visit initialization from $V_\phi$}
  \If{$s$ non-terminal}
    \State Sample up to $C_{\max}$ candidates $C_t$ from $\pi^{(i_t)}(\cdot\mid x,H_{t-1}^{(i_t)})$ where $s=(x,H_{t-1},i_t)$
    \For{each $a \in C_t$}
      \State $s' \gets \mathrm{next}(s,a)$
      \State $N(s,a)\!\gets\!1$
      \State $W(s,a)\!\gets\!V_\phi(s')$
      \State $\widehat Q(s,a)\!\gets\!V_\phi(s')$
    \EndFor
  \EndIf
  \Comment{Leaf evaluation by bootstrap}
  \State $v \gets V_\phi(s)$
  \Comment{Backpropagation}
  \For{each $(u,b)$ in $\text{path}$}
    \State $N(u,b)\!\gets\!N(u,b)+1$
    \State $W(u,b)\!\gets\!W(u,b)+v$
    \State $\widehat Q(u,b)\!\gets\!\frac{W(u,b)}{N(u,b)}$
  \EndFor
\EndFor
\State Decode: From $s_1(x)$, greedily follow at each depth the child with maximal $\widehat Q$
\end{algorithmic}
\end{algorithm*}

\begin{algorithm*}[t]
\caption{Step-Level Beam Search (SBS$(B_2,B_1)$) with \textsc{MASPRM} or Policy Score}
\label{alg:sbs}
\begin{algorithmic}[1]
\Require Instance $x$, graph $G$, schedule $\sigma$, global beam $B_1$, samples per state $B_2$, scorer $f$ where $f(s,a,s')=V_\phi(s')$ or $\tilde\psi_{\text{pol}}(a\mid s)$
\State Initialize beam $\mathcal{B} \gets \{s_1(x)\}$
\For{$t=1$ to $D$} \Comment{$D$ is the schedule length (one agent per depth, $i_t=\sigma(t)$)}
  \State $\mathcal{P} \gets [\,]$
  \For{each $s=(x,H_{t-1},i_t) \in \mathcal{B}$}
    \State Sample $B_2$ candidate actions $\{a^{(b)}\}_{b=1}^{B_2} \sim \pi^{(i_t)}(\cdot\mid x,H_{t-1}^{(i_t)})$
    \For{$b=1$ to $B_2$}
      \State $s' \gets \mathrm{next}(s,a^{(b)})$
      \State $\text{score}(s') \gets f(s,a^{(b)},s')$ \hfill (e.g., $V_\phi(s')$ or $\tilde\psi_{\text{pol}}(a^{(b)}\mid s)$)
      \State Append $(s', \text{score}(s'))$ to $\mathcal{P}$
    \EndFor
  \EndFor
  \State $\mathcal{B} \gets \text{top-}B_1 \text{ states in } \mathcal{P} \text{ by } \text{score}$ \Comment{Global pool across parents}
\EndFor
\State Select output: Return the highest scoring terminal state in the final beam
\end{algorithmic}
\end{algorithm*}

\end{document}